\documentclass[fleqn,usenatbib]{mnras}

\usepackage{newtxtext,newtxmath}

\usepackage[T1]{fontenc}
\usepackage{ae,aecompl}
\usepackage{soul}

\usepackage{graphicx}	
\usepackage{amsmath}	
\usepackage[usenames]{xcolor}
\usepackage{natbib}
\usepackage{siunitx}
\usepackage{hyperref}
\usepackage{xfrac}
\usepackage{tabularx} 

\usepackage{cases}
\usepackage{changepage}

\def\equationautorefname~#1\null{equation~(#1)}

\DeclareMathAlphabet{\mathpzc}{OT1}{pzc}{m}{it}\definecolor{purple}{RGB}{160,32,240}

\newcommand{\nick}[1]{\textcolor{black}{ #1}}

\newcommand{\Msun}{M_{\odot}}
\newcommand{\Mstar}{M_{\star}}
\newcommand{\Mearth}{m_{\oplus}}
\newcommand{\Mmerc}{m_{\rm Merc}}

\newcommand{\Rearth}{r_{\oplus}}

\newcommand{\newrev}[1]{{\color{black}#1}}
\newcommand{\red}[1]{\textcolor{black}{#1}}
\newcommand{\black}{\color{black}}

\title[Two-stage disruption]{Two-stage disruption of resonant chains 
}

\author[Choksi et al.]{Nick Choksi$^{1}$\thanks{E-mail: nchoksi@caltech.edu}, Yoram Lithwick$^{2,3}$, Eugene Chiang$^{4,5}$, and Rixin Li$^{4}$ \\ 
$^{1}$Division of Geological and Planetary Science, California Institute of Technology, Pasadena, CA 91101, USA\\
$^{2}$Department of Physics \& Astronomy, Northwestern University, Evanston, IL 60202, USA \\ 
$^{3}$Center for Interdisciplinary Exploration \& Research in Astrophysics, Evanston, IL 60202, USA \\ 
$^{4}$Department of Astronomy, Theoretical Astrophysics Center, and Center for Integrative Planetary Science, University of California, Berkeley, CA 94720, USA\\
$^{5}$Department of Earth and Planetary Science, University of California, Berkeley, CA 94720, USA \\
}

\date{Released \today}
\pubyear{2026}

\begin{document}
\label{firstpage}
\pagerange{\pageref{firstpage}--\pageref{lastpage}}
\maketitle


%
\begin{abstract} 
\red{\textit{TESS} 
is enabling the
discovery of transiting planets 
around young stars. These observations suggest that most close-in planets were born in chains of mean-motion resonances that break on a characteristic timescale of order 100 Myr.} 
This observation is surprising because the same dissipative forces 
that capture planets into resonance render their orbits long-term stable. We explore a two-stage disruption scenario for resonant chains of super-Earths. First, the chains have their (free) eccentricities excited by some mechanism. We show that any such mechanism that seeds eccentricities of a few percent sets in motion a second stage of dynamical instability on a $\sim$100 Myr timescale. A possible stage-one mechanism is the accretion of a handful of Mercury-sized bodies totaling a few percent of the planetary system mass, which excites the requisite eccentricities and triggers a stage two that reproduces the observed decline in the incidence of resonance. Impacts from such bodies can also explain why some young systems have period ratios narrow of commensurability. We sketch how these impactors may have grown out of debris left over from an earlier epoch of planet formation. We also identify two new trends in the observational data: a decline in multiplicity on the same timescale as the decline in the incidence of resonance, and an increase in the occupation of resonances with multiplicity.
\end{abstract}

\black

\begin{keywords}
planets and satellites: dynamical evolution and stability – planets and satellites: formation
\end{keywords}


\section{Introduction}
At orbital distances $\lesssim$1 au, ``super-Earths'' are the most common kind of planet \citep{fressin_etal_2013, dressing_charbonneau_2015, petigura_etal_2018,
zhu_etal_2018}. Close-in planetary systems are packed. \red{They usually host at least a few planets, spaced apart such that their orbital period ratios are  $\lesssim$3 (\citealt{ lissauer_etal_2011, fabrycky_etal_2014, sandford_etal_2019}).}

\begin{figure} 
\includegraphics[width=\columnwidth]{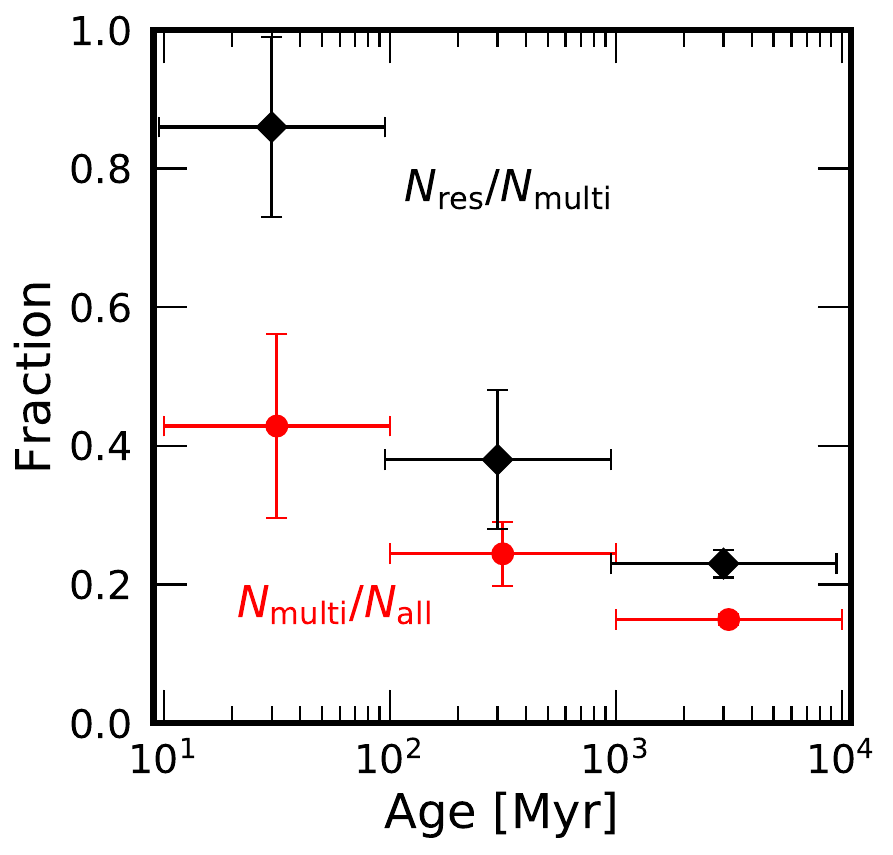}
\caption{Observed resonance and transit multiplicity statistics vs. age. Black diamonds 
reproduce the incidence of resonance reported by \protect\cite{dai_etal_2024}. This quantity is defined as the fraction of multi-transiting systems that host at least one pair near resonance. Horizontal errorbars represent the bin width and vertical errorbars represent Poisson uncertainty. As systems age, resonances become less common. \red{Red circles show the fraction of systems with multiple planets. These points are calculated using all systems that host at least one transiting planet with orbital period $<1$ yr and a stellar age listed in the NASA Exoplanet Archive.} The parallel decline in the two sets of points seems consistent with a picture in which resonances break and trigger dynamical instability, reducing observed multiplicities through mergers and excited mutual inclinations. }
\label{fig:age_multiplicity}
\end{figure}

Until recently, our understanding of these inner systems was limited to those around the Gyrs-old stars preferentially observed by the \textit{Kepler} spacecraft. That situation is changing. Thanks to its all-sky coverage, the \textit{Transiting Exoplanet Survey Satellite} (\textit{TESS}; \citealt[][]{ricker_etal_2014}) is beginning to detect close-in planets around young stars \citep[e.g.][]{newton_etal_2019, mann_etal_2020, newton_etal_2021, thao_etal_2024, dattilo_etal_2025}.
\red{
From these detections, there are indications that mean-motion resonance is common among newly formed planets.
\cite{dai_etal_2024} compiled the available data and found that at ages $\lesssim$100 Myr, $\sim$80\% of multi-transiting systems host a pair of planets near resonance. Over time, resonances disappear. For systems between 100 Myr -- 1 Gyr in age, the incidence of resonance declines to $\sim$40\%, and at $\gtrsim$1 Gyr it has fallen to $\sim$20\%. Figure \ref{fig:age_multiplicity} plots this reported trend. Qualitative support for this picture is also found in earlier \textit{Kepler} data \citep{hamer_schlaufman_2024}, and in the higher occurrence rate of transit timing variations (TTVs) among young planets \citep{lopez-murillo_etal_2026}. We
acknowledge that the ranks of young planets are still thin -- there are currently just eight transiting multiplanet systems aged less than 100 Myr.\footnote{\red{\cite{hansen_etal_2025} express skepticism in the conclusions of \cite{dai_etal_2024} because the latter include nontransiting planets detected by \textit{Kepler} and \textit{TESS} via TTVs (transit timing variations) induced on transiting planets. While resonant systems have outsized TTVs and therefore counting TTV-inferred planets biases resonant fractions upward, this bias cannot reverse the reported decline in the incidence of resonance. This is because  \textit{TESS}, with its shorter time baseline, is worse at detecting TTVs than \textit{Kepler}. Thus the resonant fraction would be more severely overestimated at the old ages probed by \textit{Kepler}, which only reinforces the claimed decline. 
In any case, we repeated the analysis of \cite{dai_etal_2024} for systems aged $<100 $ Myr, excluding the planets AU Mic d and TOI 1227 c which were discovered through TTVs. The adjusted fraction of young resonant systems drops from 7/8 to 5/6, still well above that for Gyr-old systems.}} Nevertheless, we will take these measurements at face value and interpret what appears to be a decline in the incidence of resonance on a characteristic timescale of order 100 Myr. }


A ubiquity of resonance at young ages 
is a smoking gun for planets having migrated significantly\footnote{By ``significant,'' we mean that super-Earths migrated by factors of a few in orbital radius. Larger changes seem unlikely to us given the mounting evidence that most close-in planets are not 
composed of material from beyond the water-ice line at a few au \citep[][]{rogers_owen_2021, rogers_etal_2023, dainese_albrecht_2025}.} through their parent gas disks. Migration is effected by the excitation of waves by planets in their natal gas disks, a.k.a. dynamical
friction \citep{goldreich_tremaine_1980}. As planets change their semimajor axes and their orbital period ratios approach a commensurability like 3:2 or 2:1, they can capture into resonance \citep[][]{borderies_goldreich_1984, batygin_2015}. Once captured, migrating planets exchange angular momentum at just the right rate to maintain period ratios near commensurability \citep[][]{goldreich_1965}. The equilibrium spacing places the planets slightly wide of exact commensurability \citep{choksi_chiang_2020}, as reflected in the asymmetries around resonances in a histogram of period ratios  \citep{lissauer_etal_2011, fabrycky_etal_2014}.

When resonances break, chaos ensues \citep[e.g.][]{izidoro_etal_2017}. Thus the disappearance of resonances on a $\sim$100 Myr timescale points to an extended epoch of dynamical upheaval.
Because super-Earths orbit so close to their host stars, chaos ends in collisions. Evidence for a violent era of impacts comes from the higher masses of nonresonant planets interpreted to be collisional merger products 
\citep{leleu_etal_2024,li_etal_2025}, non-zero phases of transit timing variations \citep{lithwick_etal_2012, choksi_chiang_2023}, intra-system uniformity \citep{goldberg_etal_2022}, and the smooth distribution of interplanetary spacings in mature systems \citep{izidoro_etal_2017, izidoro_etal_2019, li_etal_2025}. \red{In this scenario we would also expect transit multiplicity to decline with age as dynamical instability merges planets and excites mutual inclinations. Such a trend has been reported 
by \citet[][their fig. 4]{yang_etal_2023} among $\gtrsim$ Gyr old stars age-dated using stellar kinematics. 
In Figure \ref{fig:age_multiplicity} we extend this result by including all transiting systems with host star ages (derived from any method) in the NASA Exoplanet Archive. Our figure demonstrates that transit multiplicity declines on the same 
timescale as the incidence of resonance. The short 27 day baseline of \textit{TESS} disfavors detections of multiple planets, so the intrinsic trend may be even stronger than plotted.}



What remains unclear is why resonances break 
in the first place. The problem is that migration is a double-edged sword. Like friction on a pendulum, dynamical friction damps planetary systems to stable fixed points \citep[e.g.][]{goldreich_1965}. The resonances
that convergent migration
forms are usually so stable that they do not break within system lifetimes \citep[e.g.][]{nagpal_etal_2024, li_etal_2025}. 
Resonances can break if planets have their eccentricities excited -- technically their free eccentricities, the components of eccentricity not forced by the resonances.
\red{Most theories for resonance breaking so far invoke an exterior reservoir of other bodies, themselves eccentric enough to perturb the inner chain. These bodies could be planets or planetary embryos 
at a few au
\citep{goldberg_etal_2025, ogihara_etal_2026, lorusso_etal_2026}, or they could be planetesimals from even greater distances that are flung inward by massive companions \citep{li_etal_2025b}.} 
\color{black}
For certain choices of model parameters, these studies report disruption on a $\sim$100 Myr timescale as observed. The reasons for the long timescale behavior seen in these ex-situ scenarios are 
not entirely clear,\footnote{\cite{ogihara_etal_2026} attribute the slow disruption of their chains to the long Laplace-Lagrange forcing timescales of exterior embryos. But in some limited $N$-body experiments we found that proximity to resonance strongly shields planets from secular forcing. We suspect that many of their chains are actually broken by the subset of embryos excited to high enough eccentricity to cross orbits with the super-Earths.} but seem related to how long it takes the outer system, with its inherently longer dynamical times, to excite to high eccentricity (see fig.~15 of \citealt{goldberg_etal_2025}).

In this paper we explore a different picture. We still rely on smaller bodies to break super-Earths out of resonance. But rather than importing them from afar we consider a locally sourced population. We imagine that the end stages of close-in planet 
formation included a final phase of in-situ ``clean up'' of remnant planetesimals. In $\ll$100 Myr, the planets clear this debris and in the process are nudged off their resonant equilibria. We show by direct $N$-body integration that the perturbations so seeded can gradually grow and cause a resonant chain to destabilize on $\sim$100 Myr timescales.

Recently, \cite{hadden_wu_2026} 
proposed the same scenario, which they dubbed ``rattle and break'' for its two stages of stirring and subsequent instability. Our work complements theirs in a few ways. First, we show that the two stages are largely independent of each other: 
resonant chains of super-Earths excited to free 
eccentricities of a few percent by any mechanism, not necessarily accretion of small bodies, are shown to be generically unstable on long timescales. Second, we offer a larger suite of simulations, which allows us to survey the parameter space more broadly, and gives us the statistics to make head-to-head comparisons against the data.

Section \ref{sec:fiducial} describes the setup and evolution of a fiducial model.
Section \ref{sec:param} surveys how results change across the parameter space.
Section \ref{sec:summary} summarizes and discusses origin scenarios for the small bodies. There, we also compare our results to those of \cite{hadden_wu_2026} and explain why we reach some different conclusions.

\begin{table*}
\begin{tabular}{ccc}
\hline 
\\[-2mm]
Parameter Name &  Description & Fiducial\\ 
\hline 
$N_{\rm big}$ & Number of big bodies  & 5   \\ 
$m_{\rm big}$ & Masses of big bodies  & 3-5 $\Mearth$ \\ 
$j$ &  Resonance index of big bodies, $P_{i+1}/P_i \approx j/(j-1)$ & 3  \\ 
$M_{\rm small}$/$M_{\rm big}$ & Total mass in small bodies, normalized by total mass in big bodies & 5\%   \\ 
$m_{\rm small}$ &  Individual masses of small bodies   & 3$\Mmerc$ = 0.15$\Mearth$ \\ 
\\[-2mm]
\hline 
\hline 
\end{tabular}
\caption{Summary of model parameters and their fiducial values adopted in Section \ref{sec:fiducial}. Alternate parameters are considered in Section \ref{sec:param}.}
\label{tab:params}
\end{table*}

 \begin{figure} 
\includegraphics[width=0.95\columnwidth]{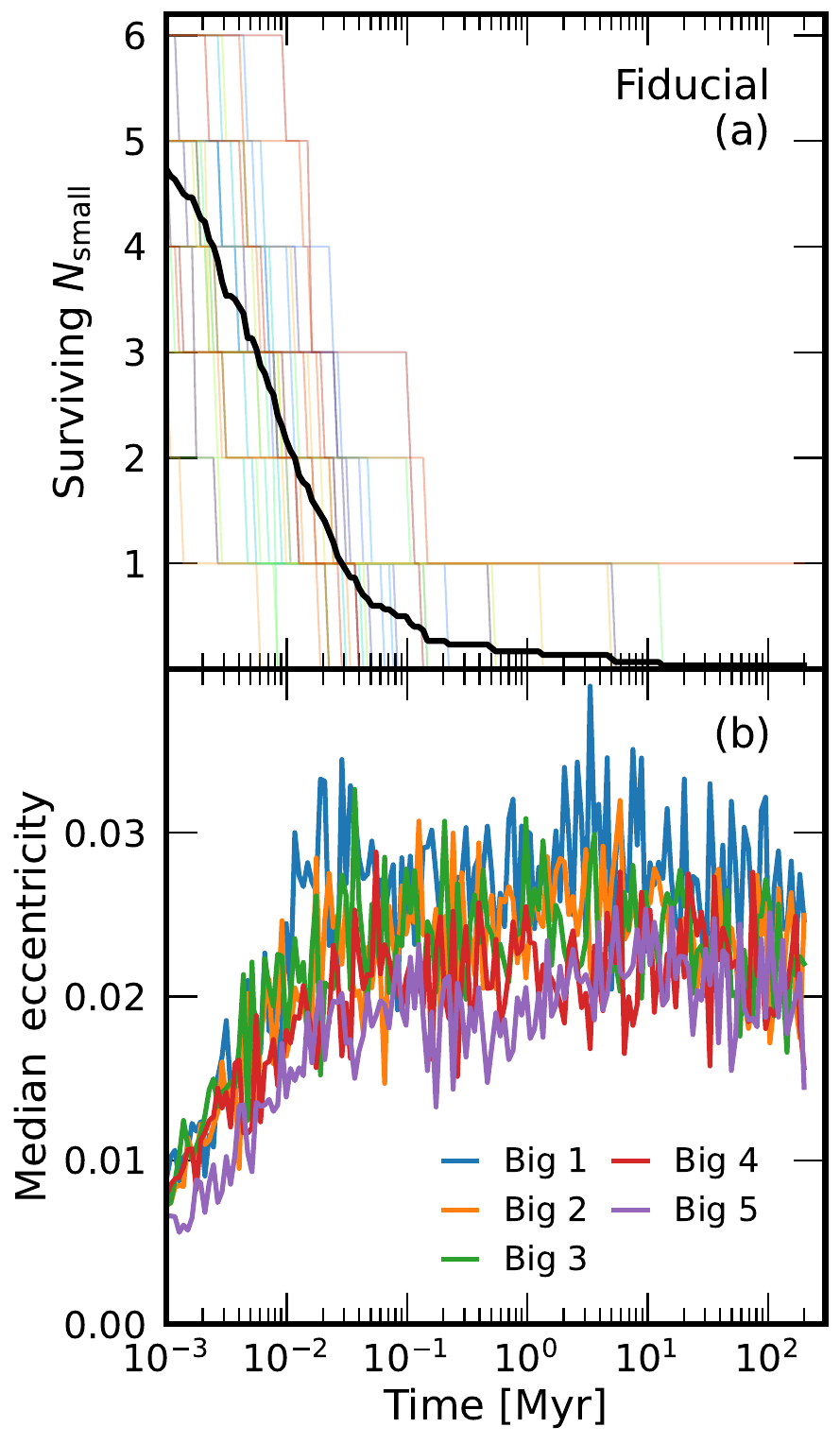}
\caption{\textit{Top}: Surviving number of small bodies vs.~time in our fiducial model. Thin colored curves plot tracks from individual chaotic realizations and the thick black curve plots their average. Most small bodies are accreted in $\lesssim 10^4\,\rm yr$.
\textit{Bottom:} Median eccentricities across all chaotic realizations for each of the five big bodies in our fiducial model. As the small bodies are accreted, they excite the eccentricities of the big bodies to a few percent.
}
\label{fig:fidsmall}
\end{figure}

\begin{figure} 
\includegraphics[width=0.95\columnwidth]{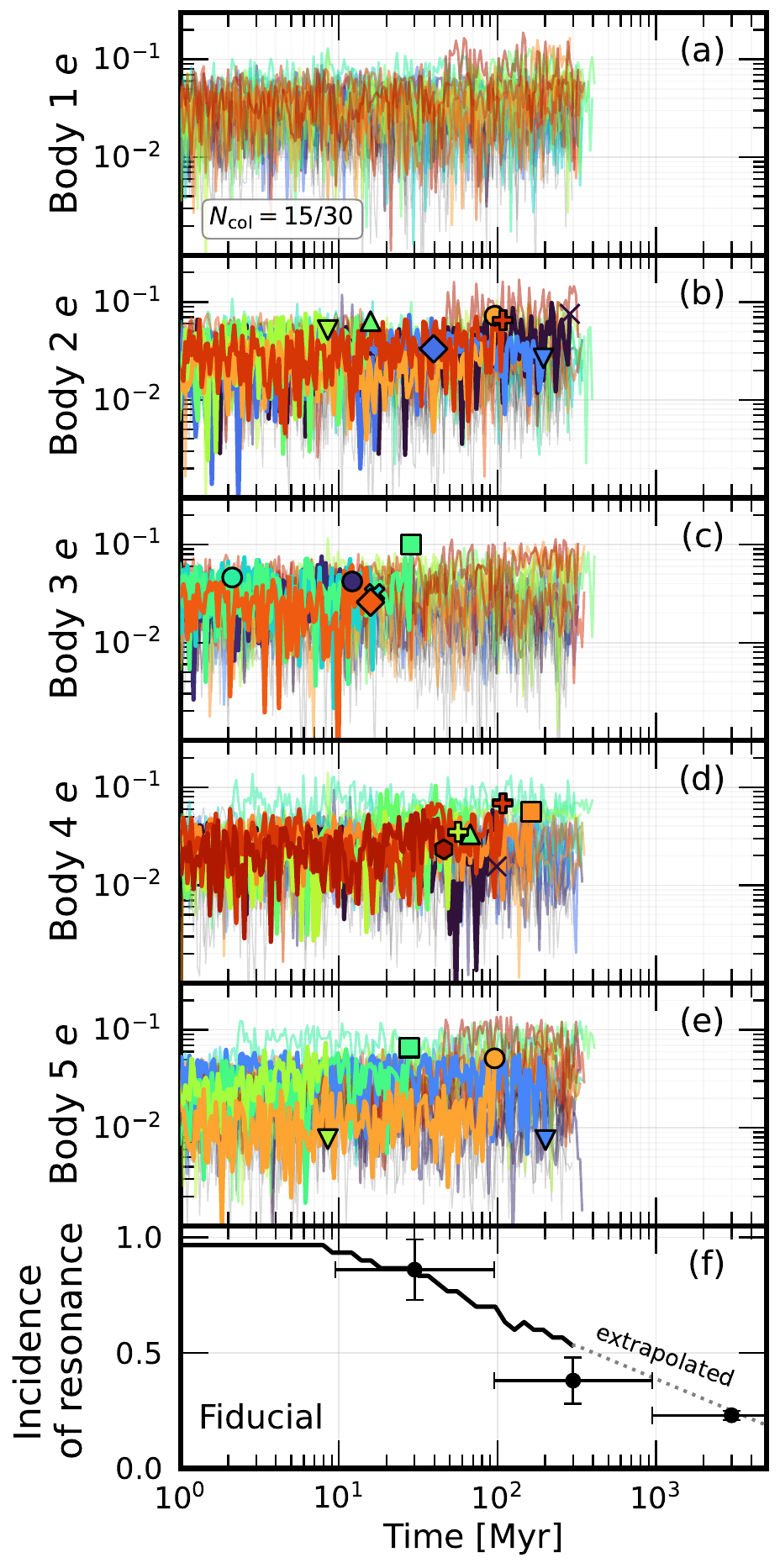}
\caption{Long-term evolution of resonant chains that have accreted a population of small bodies. The top five panels show the eccentricities vs.~time of the five big bodies in the chain. Each panel contains 30 curves, each corresponding to a different chaotic realization and assigned a different color. 
\red{After each merger between big bodies, we remove one body and update the mass of the other to equal the combined mass of the colliders. Thick curves terminated by markers represent the removed bodies.  At the end of their 300 Myr runtimes, 15 out of 30 of the simulated systems have suffered at least one collision.} 
The curve in the bottom panel plots the fraction of simulated systems with at least one pair near resonance. The dashed portion of
the curve marks a linear-log extrapolation of the simulated trend beyond the
simulation duration. Black points mark data taken from \protect \citet[][same as in Fig.~1]{dai_etal_2024}. Both the simulated and observed samples count a pair as ``near resonance'' if their period ratio $-0.015 < (j-1)P_{i+1}/(jP_i) - 1 < +0.03$.
}
\label{fig:fidbig}
\end{figure}

\section{Fiducial Model}
\label{sec:fiducial}

\subsection{Setup}
\label{subsec:setup}
Our modeled planetary systems are composed of a star (mass $1\,\Msun$), a resonant chain of super-Earths, which we call the ``big'' bodies, and a sprinkling of ``small'' bodies strewn across their orbits. Details of the setup and fiducial parameter choices are given below.

The total number of big bodies is $N_{\rm big} = 5$. Their individual masses are drawn randomly from a uniform 
distribution between $m_{\rm big} = (3-5)\,\Mearth$, where $\Mearth$ is Earth's mass. We purposefully chose these fiducial masses to be somewhat 
less 
than those of mature super-Earths weighing $\sim$7-15\,$\Mearth$ \citep{leleu_etal_2024} because in our simulations planets grow by collisions to reach the masses observed today. All big bodies have a radius $r_{\rm big} = 3\Rearth$. Neighboring big bodies have a period ratio 1\% larger than a $j$:$(j-1)$ commensurability, anchored by a fixed period of 10 days for the innermost planet. All pairs have the same $j$. For our fiducial model, we use a chain of 3:2 resonances ($j=3$) because the 3:2 is the most common resonance across all ages. The big bodies have initial eccentricities of zero and are placed randomly along their orbits. 
Initial  inclinations are drawn randomly from a uniform distribution between 0 and 1$^{\circ}$.

In initializing our resonant chains, we chose not to explicitly model an earlier phase of dissipative capture into resonance. We partially account for this by starting pairs just wide of 3:2 to mimic how capture establishes equilibrium period ratios slightly larger than perfect commensurability \citep{choksi_chiang_2020}.\footnote{\nick{In the linear theory of planet-disk interactions, a pair of $3\Mearth$ planets equilibrates 1\% wide of 3:2 if the disk aspect ratio $h \sim 0.003$ (this fractional deviation from commensurability scales as $1/h$; \citealt{terquem_papaloizou_2019}). Thus our initial condition implies a cooler disk than typically assumed. We expect that initializing pairs closer together would have little effect on outcomes. That is because scattering off small bodies quickly wedges pairs apart by a fractional amount of order $M_{\rm small}/M_{\rm big}$ \citep{wu_etal_2024, hadden_wu_2026}.}}
But because we set initial (osculating) eccentricities to zero and randomize initial orbital phases, our big bodies start with some nonzero ``free'' eccentricity (the component of their eccentricity not accounted for by gravitational interaction with neighbors). In reality, we expect disk dynamical friction to remove free eccentricity \citep{choksi_chiang_2023, goldberg_batygin_2023}. But this inconsistency matters little because the small bodies excite free eccentricities anyway. Appendix \ref{sec:appendix_damping} presents some test runs with damping and shows they behave largely the same as runs without damping.

The total mass in small bodies is $M_{\rm small} = 0.05 \times M_{\rm big}$, where $M_{\rm big}$ is the total mass in big bodies. Each small body weighs $m_{\rm small} = 3\Mmerc$, where $\Mmerc = 0.05\,\Mearth$ is Mercury's mass (section \ref{sec:summary} explores why systems might naturally breed small bodies of this mass scale). They initially number $N_{\rm small} \approx M_{\rm small}/m_{\rm small} = 4-6$. The small bodies are placed randomly along circular orbits with orbital periods drawn uniformly from 0.9 times that of the innermost big body to 1.1 times that of the outermost big body. 
Initial inclinations are drawn randomly from a uniform distribution between 0 and 1$^{\circ}$. 

 We run $N$-body simulations of these systems with \texttt{rebound} \citep{rein_liu_2012}. Our integrator is \texttt{mercurius}, which defaults to a Wisdom-Holman algorithm but switches to the adaptive-timestep \texttt{IAS15} algorithm for encounters within 3 mutual Hill radii \citep{rein_spiegel_2015, rein_tamayo_2015, rein_etal_2019}. The timestep for the Wisdom-Holman scheme is fixed to 5\% of the innermost planet's orbital period. We treat collisions as perfect mergers that conserve mass and angular momentum (see \citealt{li_etal_2025} for a study of imperfect mergers). To speed up integrations, we ignore  interactions between small bodies by modeling them as ``type I'' test particles in \texttt{rebound}.

For each parameter set, we sample the scatter in outcomes 
by simulating 30 realizations differing only in their randomly-drawn initial orbital phases. We integrate out to 300 Myr. Each realization took about 4 days of wall-clock time to run.

Table \ref{tab:params} summarizes our model parameters and their fiducial values.

 \begin{figure} 
\includegraphics[width=\columnwidth]{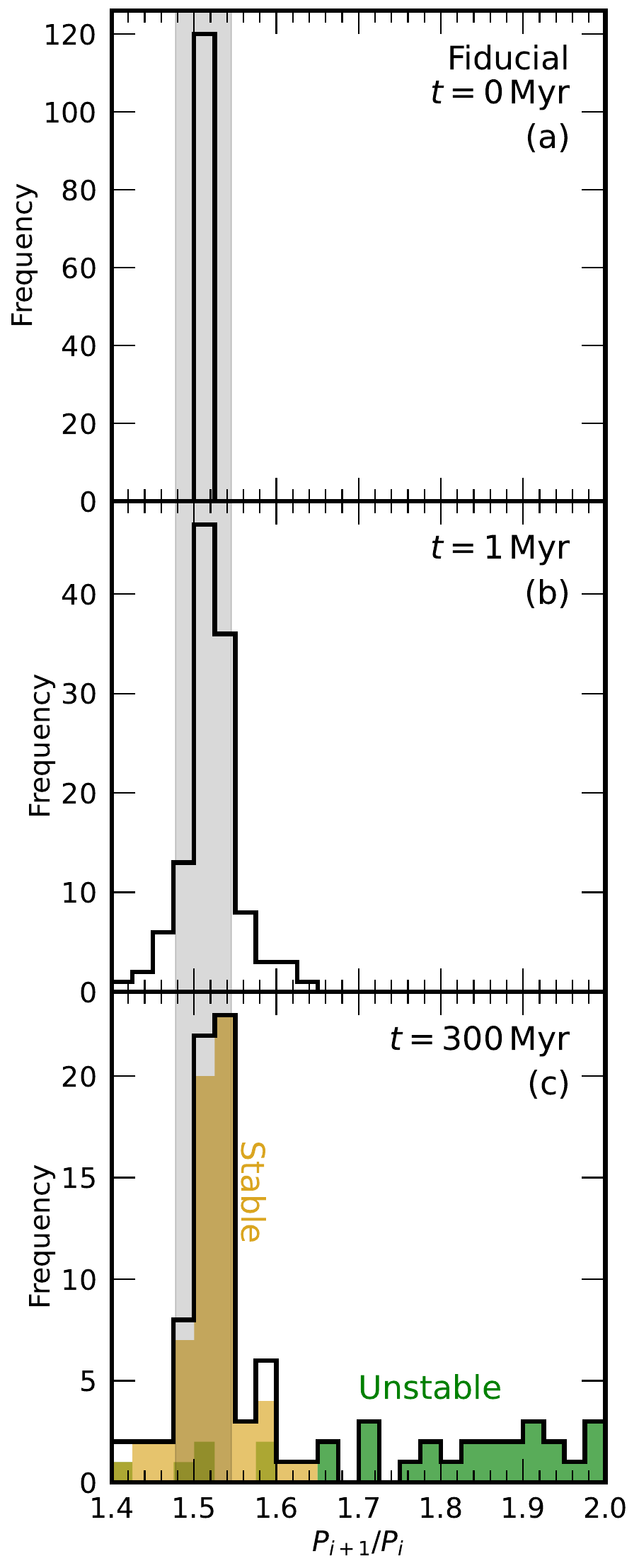}
\vspace{-0.5cm}
\caption{Distribution of period ratios for adjacent big bodies $P_{i+1}/P_i$ at three different times in our fiducial model. At $t=0$ we initialize all pairs with a period ratio of $1.515$. After $t=1$ Myr, accretion of the small bodies has smeared out this resonant peak. At $t = 300$ Myr, dynamical instabilities have eroded the peak and filled in a nonresonant continuum of period ratios. The last panel splits the distribution into a gold contribution from ``stable'' systems (defined as those that did not experience a collision) and a green contribution from ``unstable'' systems (those that experienced at least one 
collision). The grey shaded band in all panels highlights the range of period ratios we define as ``near resonance'' in Fig.~\ref{fig:fidbig}. Note the different $y-$axis scales between panels. 
}
\label{fig:fid_periodratio}
\end{figure}

 \begin{figure*} 
\includegraphics[width=\textwidth]{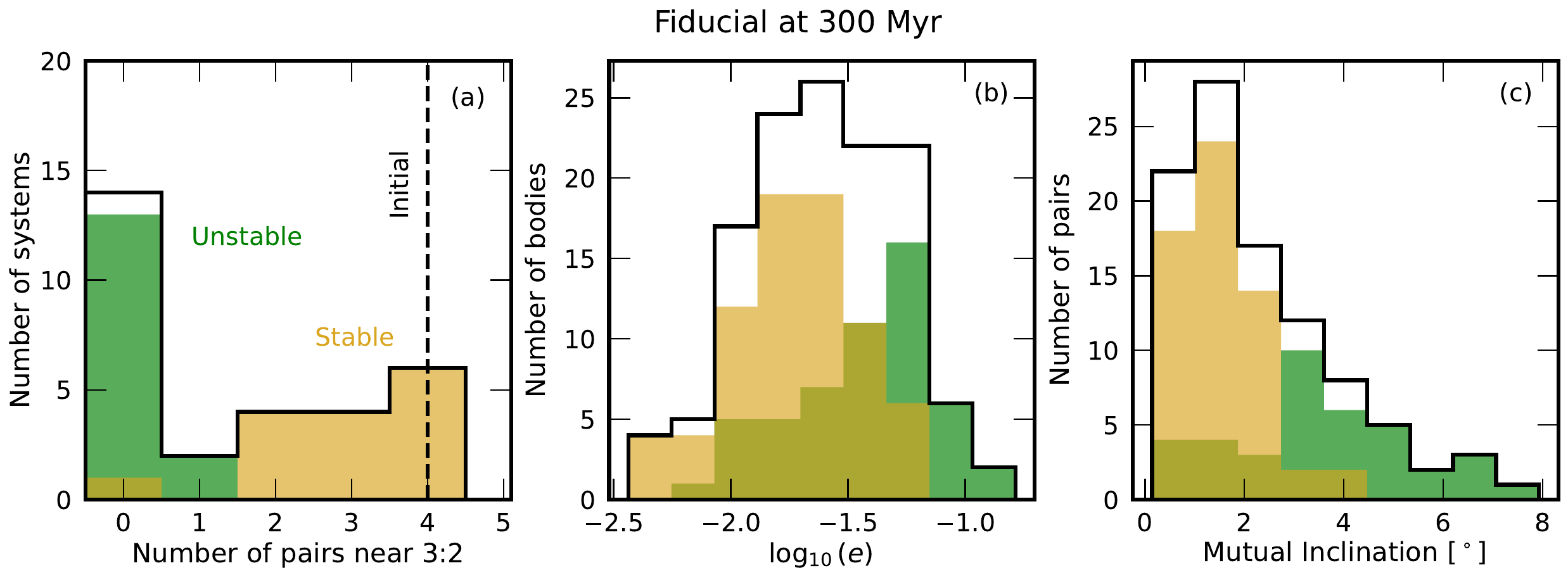}
\vspace{-0.35cm}
\caption{At the end of our simulations, unstable systems (green; those that experienced at least one collision) have different architectures from stable systems (gold; those that did not collide). The unstable systems almost always end with zero resonant pairs (left panel), larger eccentricities (middle), and larger mutual inclinations (right; measured between adjacent planets). Stable systems most often retain all four of their initial resonant pairs and are dynamically cold. In all panels, the black histogram sums the contributions from stable and unstable systems. 
}
\label{fig:nres}
\end{figure*}

\begin{figure} 
\includegraphics[width=\columnwidth]{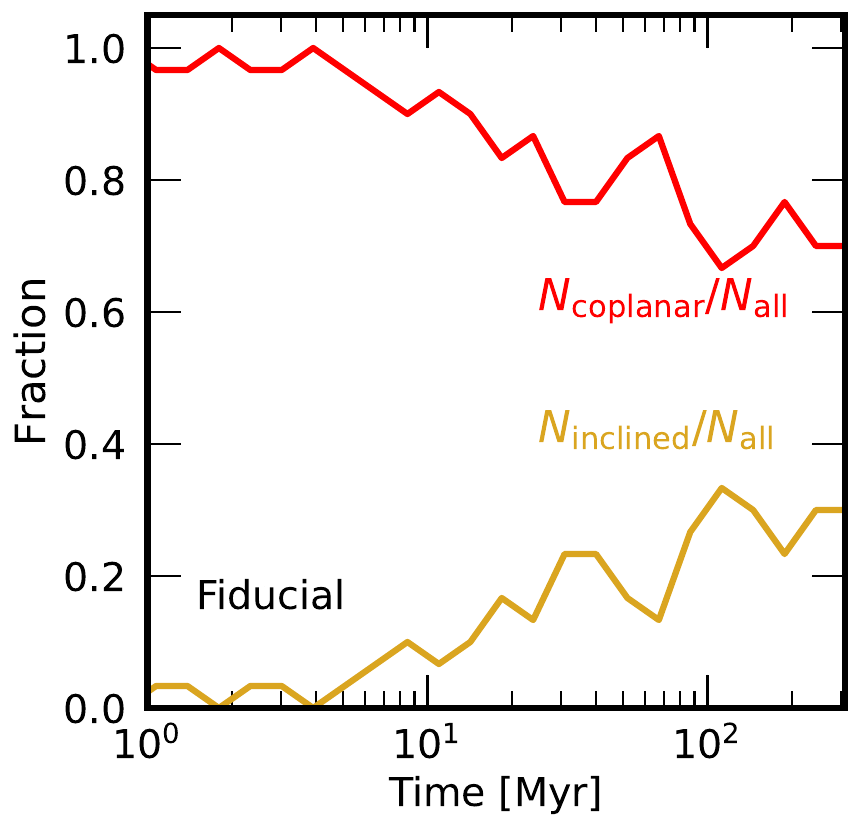}
\caption{\newrev{Fraction of simulated systems that are ``coplanar'' or ``inclined'' vs. time. Here we classify systems as coplanar if the mutual inclination between any neighboring pair of planets is below $R_{\star}/a$, where $a$ is the semimajor axis of the outer pair member. Such systems could be observed as multi-transiting, while the inclined systems would be more likely to appear as singly-transiting. In our simulations, all systems start off nearly coplanar by assumption. At $t \gtrsim 10$ Myr, dynamical instabilities start to break resonances and excite mutual inclinations. This stirring reduces the number of coplanar systems, helping to explain the observed decline in transit multiplicity plotted in Fig.~\ref{fig:age_multiplicity}. 
}}
\label{fig:fmulti_evolution}
\end{figure}
 
\subsection{Results}
In the earliest stages of our simulations, big bodies consume small bodies.
An order-of-magnitude estimate for the time it takes to accrete the small bodies is \citep[e.g.][]{goldreich_etal_2004}:
\begin{align}
t_{\rm acc} &\sim \left(\frac{a}{r_{\rm big}}\right)^2P  \nonumber  \\ 
&\sim 10^4\,\mathrm{yr}\left(\frac{r_{\rm big}}{3\,\Rearth}\right)^{-2}\left(\frac{P}{10\,\rm days}\right)^{7/3},
\label{eqn:tacc}
\end{align}
where $a$ and $P$ are orbital radius and period, and we have neglected gravitational focusing. Figure \ref{fig:fidsmall}a shows that our simulations are on average consistent with this estimate. 
Accretion and scattering of the small bodies excite the eccentricities of the big bodies to a few percent (Fig.~\ref{fig:fidsmall}b).
That these values are much greater than resonantly forced eccentricities $\mathcal{O}(10^{-3})$ (e.g. equation A15 of \citealt{lithwick_etal_2012} or our Fig.~\ref{fig:damp_big}) supports our choice not to model dissipative capture into resonance (see also appendix \ref{sec:appendix_damping} for explicit tests).

After this early stage of accretion and stirring, some of our simulated resonant chains suffer a dynamical instability.  We define ``unstable'' systems as those that experienced at least one collision, and
``stable'' systems as those that did not. Figure \ref{fig:fidbig} plots eccentricity tracks for individual chaotic realizations and highlights systems that go unstable. Within our simulation runtime of 300 Myr, about half of systems 
suffer an instability. Most collisions happen on timescales of 10-100 Myr, long after all the small bodies have been cleared.

The bottom panel of Figure \ref{fig:fidbig} compares our simulated systems against the data. We calculate the ``incidence of resonance,'' defined by \cite{dai_etal_2024} as the fraction of systems hosting at least one planet pair with $-0.015 < (j-1)P_{i+1}/(jP_i) - 1 < +0.03$. In our simulations, the fraction is at 100\% at $t=0$ by assumption. It remains constant for the first 10 Myr or so, and then starts to decline. At $t = 30$ Myr, it has reached 80\%, and at $t=300$ Myr it has fallen further to 50\%. \newrev{Our simulations end at this point. Without further integration, we cannot say with certainty how our modeled systems will evolve. But we see no indication in Fig.~\ref{fig:fidbig}f that instabilities are coming to a halt. From the available simulation data, 
the number of disrupted systems is roughly constant per logarithmic time interval. Such behavior  
has been reported in numerical studies of dynamical instability in nonresonant systems (fig.~7 of \citealt{holman_wisdom_1993}; \citealt{pu_wu_2015}). We therefore extrapolate our results by fitting a line to the simulated $\log t$ vs. incidence of resonance trend for $t > 10$ Myr. Projecting one decade forward in time suggests the incidence of resonance will reach $\sim$20\% at $t =$ 3 Gyr. Longer-term integrations to test this  extrapolation would be welcome.}



Figure \ref{fig:fid_periodratio} plots histograms of period ratios for neighboring big bodies. Our initial conditions at $t=0$ start all pairs with period ratios $P_{i+1}/P_{i} = 1.515$ (Fig. \ref{fig:fid_periodratio}a). At $t = 1$ Myr, after all the small bodies have been accreted, the peak has widened to include somewhat larger period ratios 
(Fig. \ref{fig:fid_periodratio}b). 
On the flip side, we also find a handful of systems with period ratios $P_{i+1}/P_i < 3/2$. Young planet pairs narrow of resonance are observed (see the compilation in \citealt{dai_etal_2024}), and are not a predicted outcome of convergent migration acting alone \citep{choksi_chiang_2020}. At the end of our simulations, a broad continuum of period ratios has begun to
fill out (Fig.~\ref{fig:fid_periodratio}c). The continuum is made up of planet pairs that were once near resonance but suffered dynamical instability (green). 
Collisions widen interplanetary spacings, so breaking our fiducial 3:2 resonant chains fills out the continuum between 3:2 and 2:1 (see also \citealt{li_etal_2025}).

Our simulated chains exhibit a dichotomy in outcomes. Unstable systems practically always end up dramatically reorganized (green histogram in Figure~\ref{fig:nres}a). Stable systems, on the other hand, most commonly retain all planets near resonance. Fig.~\ref{fig:nres}bc show that the planets in unstable systems have elevated mutual inclinations of zero to eight degrees and eccentricities of a few to ten percent. Such values are consistent with strong scattering between the big bodies, which excites inclinations and eccentricities of order $v_{\rm esc}/v_{\rm K} \sim 0.1$, where $v_{\rm esc}$ is the surface escape speed from the planet and $v_{\rm K}$ is the local orbital speed. That the excited inclinations straddle $R_{\star}/a \sim 3^{\circ}$ ($R_{\star}$ is the star's radius), the value above which some bodies may be missed by transit surveys, seems qualitatively consistent with the observed decline in the frequency of multi-transiting systems shown in Fig.~\ref{fig:age_multiplicity}. 

\newrev{We cannot directly compare to the data in Fig.~\ref{fig:age_multiplicity}
without mock observing our simulated systems \citep[e.g.][]{hansen_murray_2013, he_etal_2019}. As a crude proxy, we classify systems as either ``coplanar'' or ``inclined''. Coplanar systems are defined as those in which at least one pair of planets has a mutual inclination less than $R_{\star}/a$, taking $a$ as the larger of the two semimajor axes; such systems have a chance of being observed as multi-transiting. Figure \ref{fig:fmulti_evolution} shows that the fraction of coplanar systems in the simulations has declined by about 30\% at $t=300$ Myr. This simulated trend can account for some of the evolution of observed systems: between the 0-100 Myr and 100 Myr - 1 Gyr age bins, the fraction of observed systems with multiple planets drops by a factor of 2.}



\begin{figure} 
\includegraphics[width=\columnwidth]{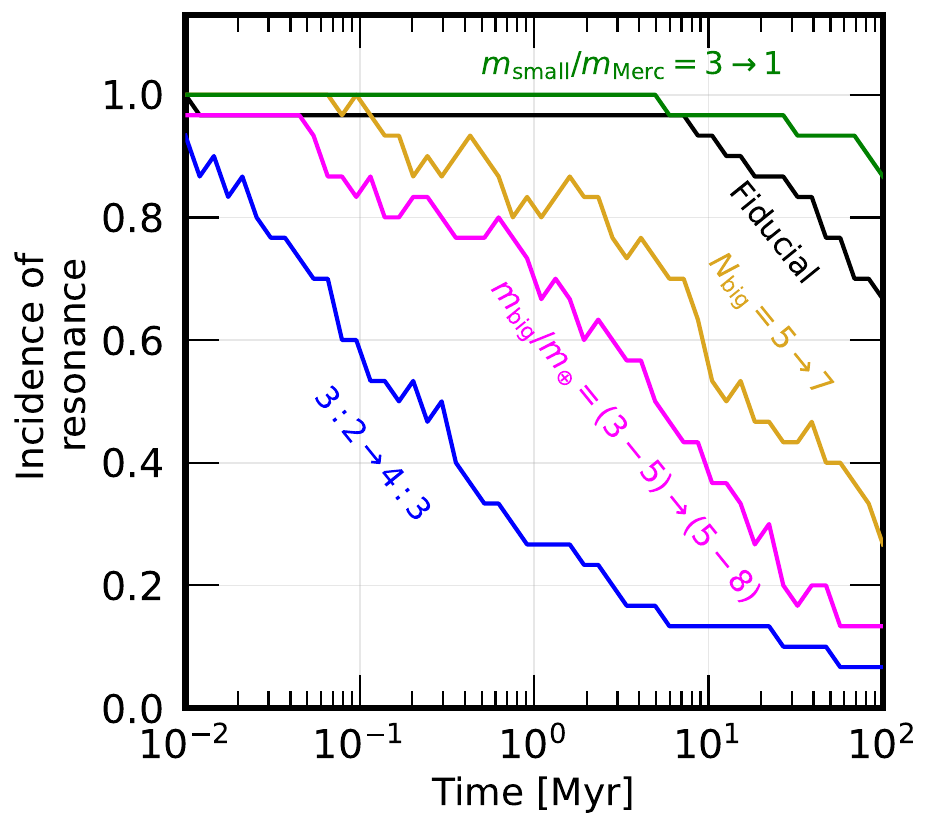}
\vspace{-0.4cm}
\caption{How different parameters affect resonance stability. Each curve plots the fraction of 30 simulated systems with at least one pair near resonance. Annotations describe the changes relative to the fiducial model, whose parameters are also summarized in Table \ref{tab:params}. \red{The 
shorter timescale jitter in some curves is due to pairs being nudged into and out of our threshold for near-resonance.
} } 
\label{fig:survey}
\end{figure}

 \begin{figure} 
\includegraphics[width=0.95\columnwidth]{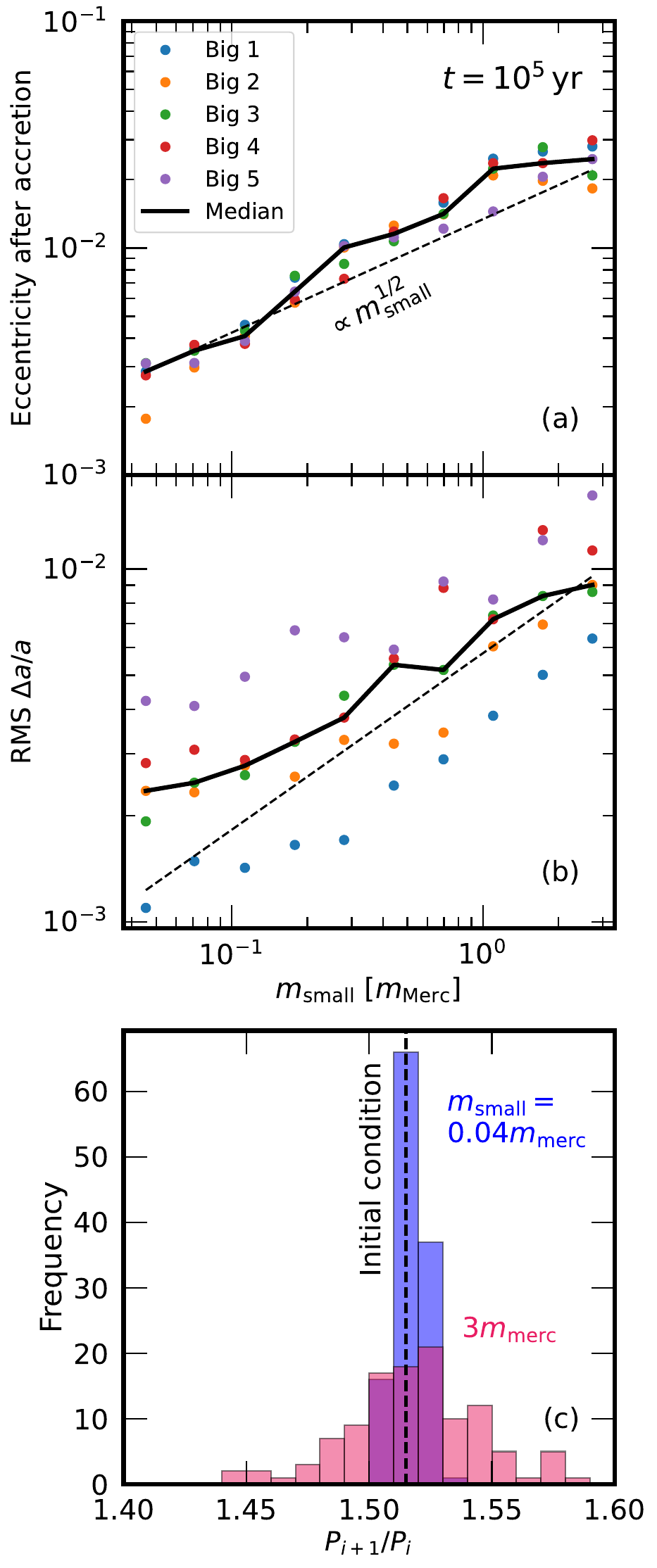}
\vspace{-0.4cm}
\caption{How accretion of small bodies affects the orbits of big bodies.
The top panel shows the median eccentricity of the five big bodies for different small body masses $m_{\rm small}$, at fixed total mass $M_{\rm small}$. Data are recorded from the simulation at $t=10^5$ yr, after most small bodies have been accreted but before instability sets in. The thick solid curve marks the median eccentricity of all the big bodies. 
The middle panel plots the root-mean-square change in semimajor axis $\Delta a/a$. The dashed lines in both panels plot the $m_{\rm small}^{1/2}$ scaling expected of a random walk, adjusted in normalization to match the median numerical results. \red{Larger fractional semimajor axis changes for the outer planets reflect their fractionally larger changes in energy  from small-body impacts.} The bottom panel contrasts the period ratio distribution after accretion for two different values of $m_{\rm small}$. Concentrating the mass in a fewer number of small bodies boosts stochasticity, which smears out the resonant peak. The vertical dashed line marks the initial condition for all pairs $P_{i+1}/P_i = 1.515$. }
\label{fig:ecc_msmall}
\end{figure}

\section{Alternate models }
\label{sec:param}
\subsection{Parameter survey}
\label{subsec:survey}
We start by changing one parameter at a time from our fiducial model. We still simulate 30 chaotic realizations for each parameter set, but reduce the runtime to 100 Myr. Figure \ref{fig:survey} summarizes the results. The most dramatic effects on stability come from changing the properties of the big bodies: 
the compactness of their resonances (the value of $j$), the number of bodies in the chain, and their masses. Switching from a series of 3:2 resonances to a series of 4:3 resonances shortens the time to instability by a factor of $\sim$10$^3$; increasing planet masses by 50\% shortens it by a factor of 30; and lengthening the chain from 5 to 7 planets shortens it by a factor of 10.

Figure \ref{fig:survey} also shows that concentrating the perturbing mass into fewer bodies promotes instability. 
That is because a handful of more massive perturbers are better at exciting eccentricity than a sea of small ones. To illustrate this, we ran a  batch of simulations with varying $m_{\rm small}$. We set the runtime of these simulations to 10$^5$ years, long enough that the small bodies have been consumed, but short enough that instability has not yet set in. Figure \ref{fig:ecc_msmall}a shows that the post-accretion eccentricity grows as $m_{\rm small}^{1/2}$. To reproduce this scaling, we consider a small body on an eccentric orbit impacting a big body on a circular orbit. The former has a typical radial speed $u_r \sim e_{\rm small}v_{\rm K}$, where $e_{\rm small}$ is its eccentricity, $v_{\rm K} = \sqrt{G\Mstar/a}$, and both bodies have about the same semimajor axis. From momentum conservation the planet's radial speed grows to $v_r \sim u_r \times m_{\rm small}/m_{\rm big}$, corresponding to an eccentricity $ e  \sim v_r/v_{\rm K}$. For a series of randomly oriented impacts, the eccentricity random walks. Assuming each big body consumes $\sim N_{\rm small}/N_{\rm big}$ small bodies, the eccentricity grows to
\begin{align}
e_{\rm acc} &\sim F\frac{m_{\rm small}}{m_{\rm big}}\, e_{\rm small}
\sqrt{N_{\rm small}/N_{\rm big}} \nonumber \\ 
&\sim 0.02\left(\frac{m_{\rm small}}{3\,\Mmerc}\right)^{1/2}\left(\frac{m_{\rm big}}{4\,\Mearth}\right)^{-1/2}\left(\frac{M_{\rm small}}{0.05M_{\rm big}}\right)^{1/2}\left(\frac{e_{\rm small}}{0.3}\right),
\label{eqn:randomwalk1}
\end{align}
where we inserted $F= 1.5$ to match the normalization of our numerical results in Fig.~\ref{fig:ecc_msmall}a.

We also consider how accretion changes a big body's orbital period. To leading order in eccentricity, the change in orbital period is controlled by the change in azimuthal velocity. The typical relative azimuthal velocity of the small body is of order its radial velocity $u_{\phi} \sim e_{\rm small}v_{\rm K}$. Thus the root-mean-square (RMS) change after many impacts is $(\Delta P/P)_{\rm acc} \sim |e_{\rm acc}|$ as given by equation \ref{eqn:randomwalk1}. Figure \ref{fig:ecc_msmall}b confirms this scaling numerically. In our fiducial model each super-Earth is hit by of order one super-Mercury. So, their orbital periods change in either direction by of order a percent, enough to smear out the distribution of period ratios around resonance as illustrated in the red histogram of Figure \ref{fig:ecc_msmall}c.

So far we have focused on the final impact and neglected the scattering that preceded it.  To weigh the relative importance of these two effects, we ran a simulation in which small bodies were removed upon impact without depositing their momentum. We found the eccentricities of the big bodies were factors of a few lower, confirming that accretion dominates over scattering in exciting eccentricities. The effect of scattering on semimajor axes is more subtle. Scattering 
causes resonant pairs to wedge apart \citep{wu_etal_2024, hadden_wu_2026}. In the distribution of period ratios, this systematically shifts the resonant peak to larger values by an amount that depends on the total mass in small bodies. This behavior can be made out when $m_{\rm small}$ is low enough that the stochastic effects of impacts largely cancel out (see the blue histogram's modest shift rightward relative to the initial condition in Fig.~\ref{fig:ecc_msmall}). When $m_{\rm small}$ is larger, the shift is masked by the smearing out of the distribution (red histogram in Fig.~\ref{fig:ecc_msmall}).

\begin{figure*} 
\includegraphics[width=0.8\textwidth]{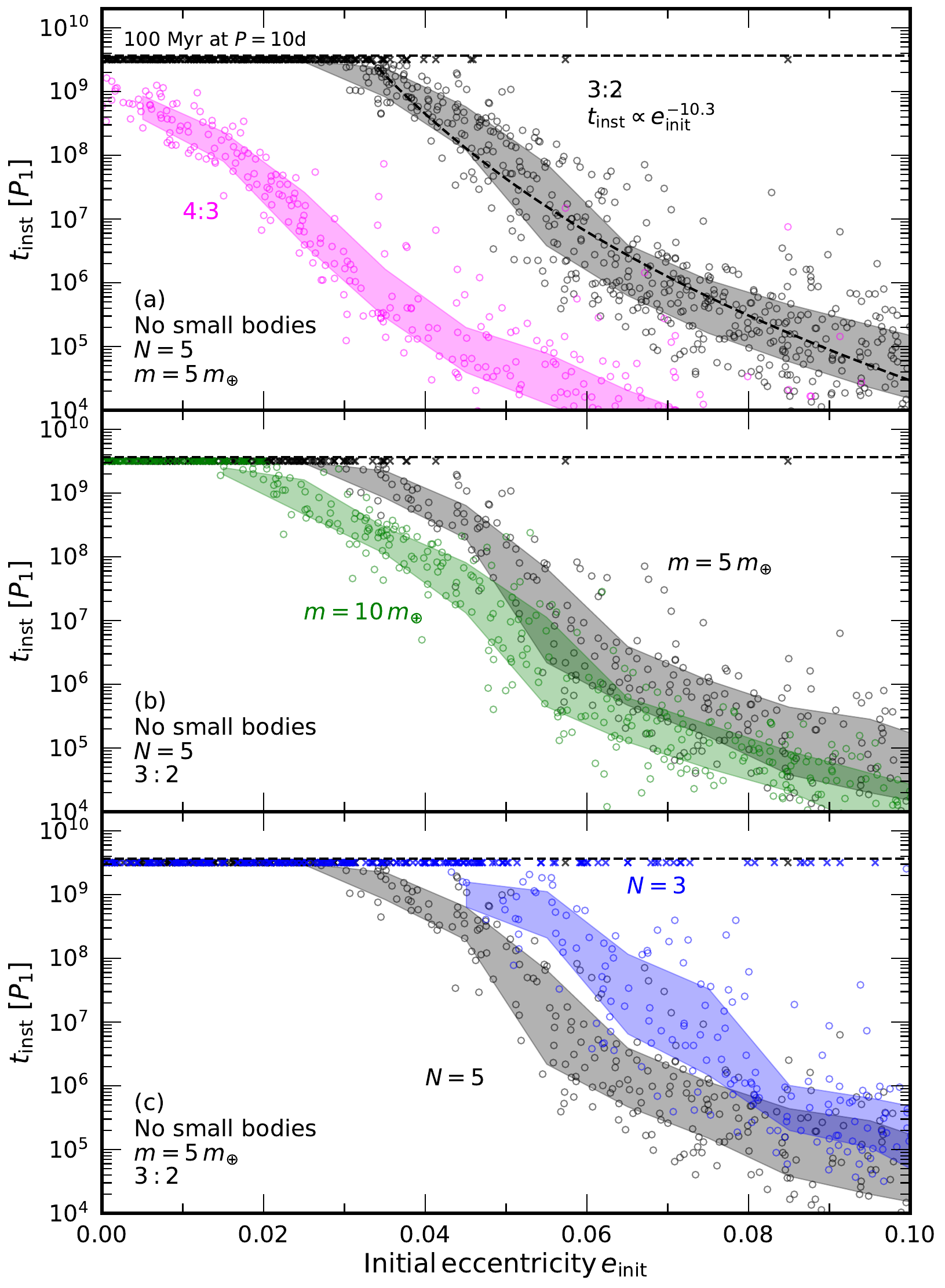}
\caption{How instability times depend on initial eccentricity. In these runs, which do not include small bodies, we initialize an $N$-planet resonant chain with nonzero eccentricities $e_{\rm init}$ as labeled. The planets start 1\% wide of either 3:2 or 4:3 commensurability with random orbital phases and have fixed planet masses of 5$\Mearth$ or 10$\Mearth$ as labeled. Open circles mark the time of the first collision, measured in units of the innermost planet's orbital period $P_1$. Shading highlights their interquartile region (25th-75th percentiles of points). Runs that did not include a collision within the simulation duration of $3 \times 10^9 P_1$ are marked with Xs. The dashed curve in the top panel shows a power-law fit to the open circles which yields $t_{\rm inst} \propto e_{\rm init}^{-10.3}$. In these runs only, we inflated the physical sizes of the planets to their Hill radii. Rerunning the 3:2 case in the top panel with $10\times$ smaller radii gave indistinguishable results.
}
\label{fig:tcol_e}
\end{figure*}

\subsection{Runs without small bodies}
\label{subsec:nosmall}
In previous sections we saw that chain breaking proceeds in two stages. First, the big bodies have their eccentricities excited. Then on much longer timescales, they undergo dynamical instability. To isolate the second stage, we run an alternate set of simulations that discards the small bodies and initializes 
the planets in the chain with some nonzero eccentricity $e_{\rm init}$ (see the caption of Figure \ref{fig:tcol_e} for more details). \red{Our goal is to understand how instability time scales with initial (free) eccentricity, 
independent of its physical origin.}

Figure \ref{fig:tcol_e} shows that the instability time $t_{\rm inst}$ (defined as the time the first collision happens) is very sensitive to initial eccentricity. The more eccentric the initial orbits, the faster instability sets in. The instability times vs.~eccentricity curves share similar shapes but shift toward faster instability for more compact resonances, higher multiplicities, and higher planet masses. Instability is also stochastic, with 1--2 orders of magnitude of scatter at fixed initial eccentricity. \red{We also tried some runs in which we first damped the eccentricities of the planets and then applied an impulsive kick to their eccentricity vectors of magnitude $e_{\rm init}$ and random direction. The resulting instability times were indistinguishable from those shown in Fig.~\ref{fig:tcol_e} (data not shown). In the Appendix, we similarly show that damping does not affect instability times in our runs with small bodies.}

We conclude that resonant chains are generically unstable on long timescales. \red{For a 3:2 resonant chain to destabilize on timescales of order 100 Myr, seed eccentricities must lie in a narrow range $e_{\rm init} \approx 2$--$5\%$.} Thus whatever mechanism excites the eccentricity needs to be finely tuned. Nonetheless, free eccentricities of resonant systems measured from transit timing variations across the age spectrum lie in or near this range \citep{hadden_lithwick_2017,hu_etal_2025}.

\section{Summary \& Discussion}
\label{sec:summary}

\begin{figure} 
\includegraphics[width=\columnwidth]{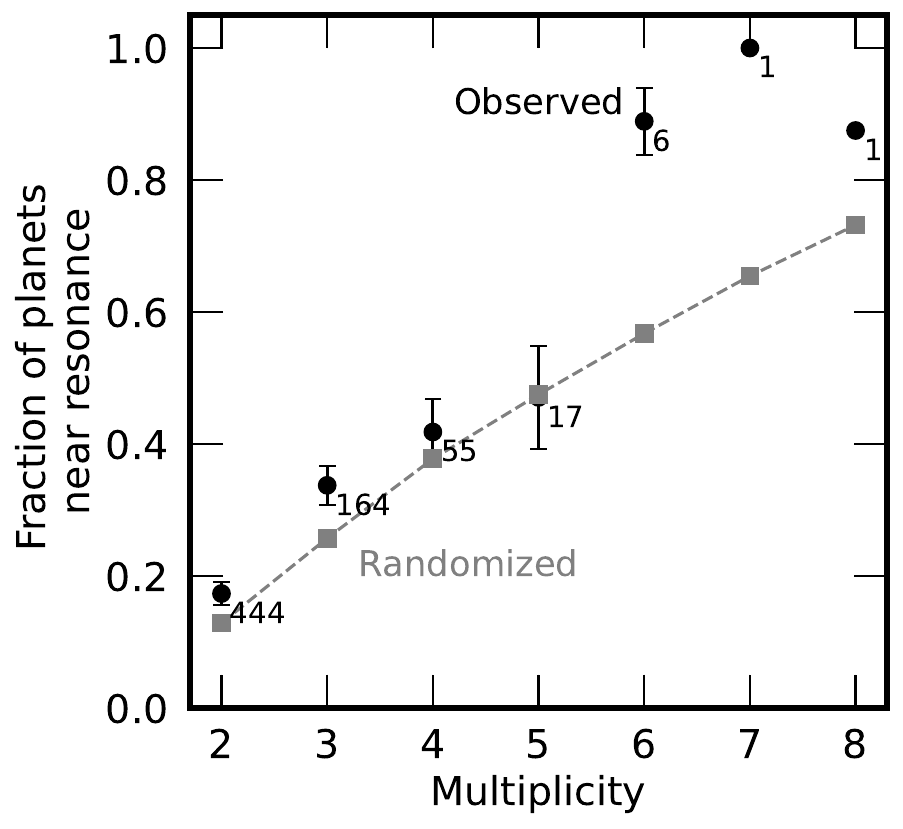}
\vspace{-0.5cm}
\caption{ \newrev{Fraction of planets found near resonance vs.~system multiplicity. Points show average values calculated from NASA Exoplanet Archive data for adjacent transiting planets. An adjacent pair of planets are counted as near resonance if they lie within $-0.015 < \Delta < +0.03$ of a first-order resonance or $-0.015 < \Delta < +0.015$ of a second order resonance, where $\Delta = (j-k)P_{i+1}/(jP_i) - 1$ is the fractional deviation from a commensurability of index $j$ and of order $k$. We also include the 8:5 using the latter threshold because it appears in TRAPPIST-1. Numbers next to the points label the count of systems at that multiplicity and errorbars represent the standard error of the mean. The $N=7$ and $N=8$ points do not have errorbars because each only represents one system (TRAPPIST-1 and Kepler-90, respectively). In high-multiplicity systems, nearly all planets are found close to resonance. But there is a sharp drop in resonance occupation for $N
\leq 5$. The grey squares show an estimate for how often we would expect to find resonances purely by chance. They come from a Monte Carlo model in which $N$ planets are drawn randomly in log-period between 10 days and 365 days, discarding systems in which two planets lie within 10 mutual Hill radii of each other. The success of the randomized model at low multiplicities, and the sudden excess above it at higher multiplicities, point to a picture in which systems formed in long resonant chains and later suffered wholesale instabilities that scrambled their orbital periods and reduced their transit multiplicities. Fig.~\ref{fig:periodratio_multiplicity} offers further observational support of this picture.
}}
\label{fig:resfrac}
\end{figure}

\begin{figure*}
\includegraphics[width=\textwidth]{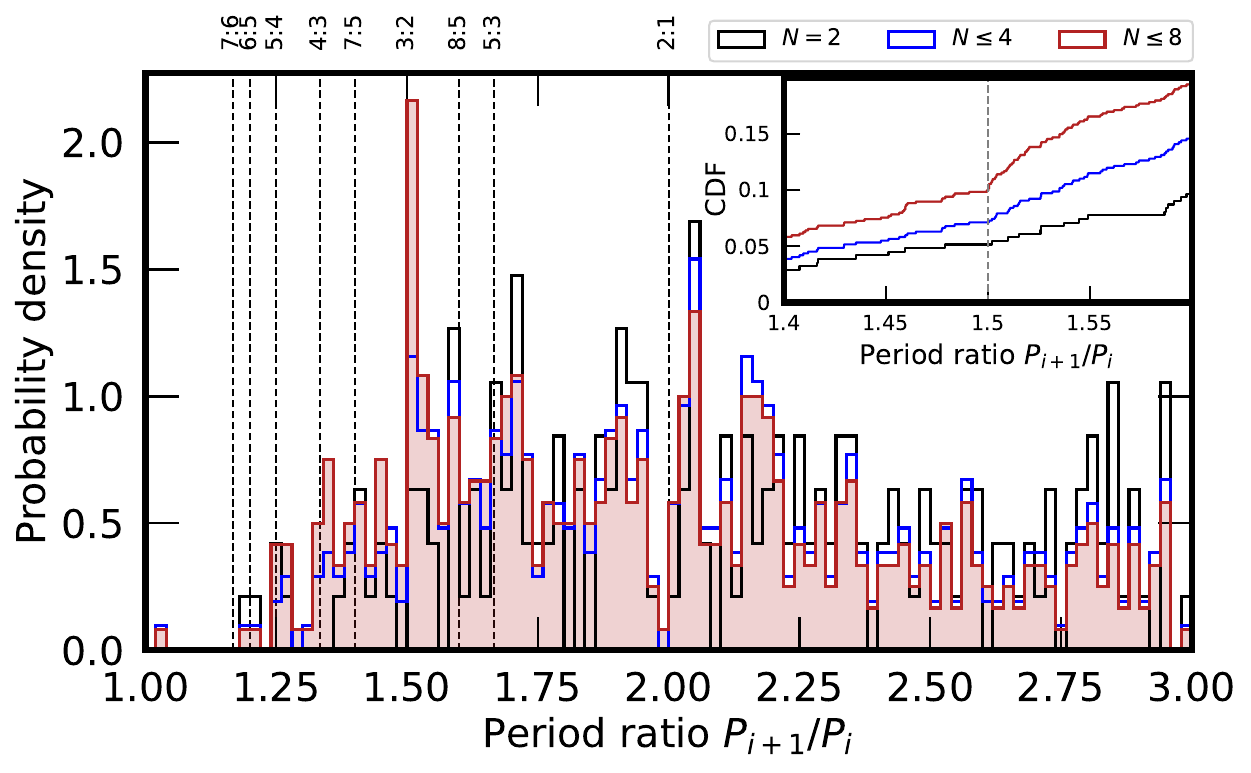}
\vspace{-0.5cm}
\caption{\newrev{Histogram of period ratios $P_{i+1}/P_i$ for all neighboring
planets discovered by transits, split by system multiplicity $N$
as labeled. Each histogram is normalized so that it integrates to unity. Data are drawn from the NASA Exoplanet Archive. A resonant 3:2 peak rises above the continuum only when
the highest-multiplicity systems are included. The inset demonstrates the same point by plotting the cumulative distribution function (CDF), zoomed-in around 3:2. Strong inclines around the resonance are only observed when high-multiplicity systems are folded in. Together with Fig.~\ref{fig:resfrac}, this result suggests that systems are
born as long resonant chains.
Most such chains undergo a wholesale
disruption that reduces system multiplicities and randomizes their orbital periods. The continuum of period ratios is made up of those disrupted systems, while the resonant peak is made up of the minority of long chains that avoid instability. 
}}
\label{fig:periodratio_multiplicity}
\end{figure*}

\begin{figure} 
\includegraphics[width=\columnwidth]{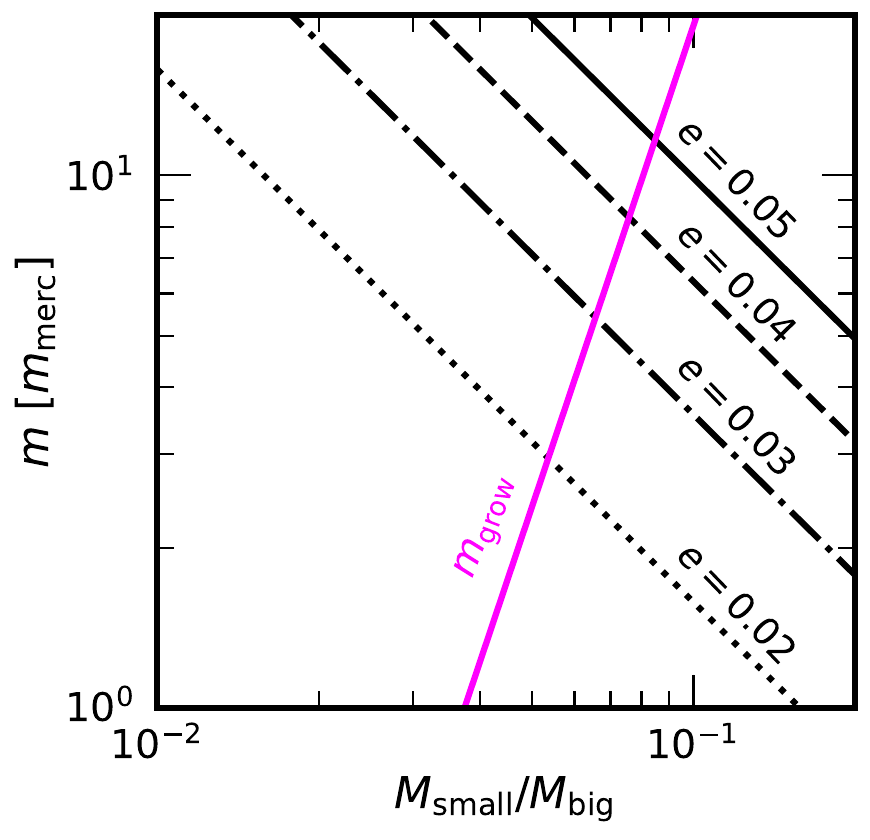}
\vspace{-0.5cm}
\caption{
The pink solid line shows the mass $m_{\rm grow}$ to which solids can grow. This value is plotted as a function of the total mass in debris $M_{\rm small}$, normalized by the total mass in super-Earths $M_{\rm big} = 20\,\Mearth$. Solid debris that starts with typical sizes anywhere below the pink line can grow up to $m_{\rm grow}$ before consumption by the super-Earths. The black lines give curves of constant eccentricity excitation (equation \ref{eqn:mrattle}). To reproduce instability on timescales of order 100 Myr, we found that eccentricities should be excited to $2-5\%$. That constrains the solid disk mass to the range $M_{\rm small}/M_{\rm big} \approx 0.05-0.09$.
}
\label{fig:grow}
\end{figure}

Resonant chains of super-Earths are observed to break apart on a timescale of $\sim$100 Myr. We explored a two-stage disruption scenario for these chains. In the first stage, which lasts $\ll 100$ Myr, the super-Earths have their eccentricities excited. Once the excitation is complete, the slower second stage commences. We showed that a chain entering stage two with eccentricities of a few percent undergoes a dynamical instability on a $\sim$100 Myr timescale, independent of how the eccentricities were excited (Fig.~\ref{fig:tcol_e}). 

For the stage-one eccentricity excitation, we considered the accretion of a handful of Mercury-mass bodies.
Our perturbers are larger than those of 
\cite{hadden_wu_2026}, who found masses of $m_{\rm Pluto} \approx \Mmerc/25$ 
sufficed to disrupt analogues of the resonant chain around HD 110067 on 
$\lesssim$100 Myr timescales. We suspect they were driven toward less massive perturbers because (i) their modeled chain includes 4:3 resonances whereas ours are composed entirely of 3:2s, (ii) their adopted planet masses range from 5-10 $\Mearth$, whereas we draw from the range 3-5$\Mearth$, and (iii) their chains have $N=6$ planets whereas ours have $N=5$. All of 
their parameter choices lower 
the critical eccentricity for stage-two instability, necessitating less massive perturbers (Fig.~\ref{fig:survey}).

Some young planetary systems show hints that they have been stirred by small bodies. The first signs are in their period ratio distribution. It is useful to define the fractional deviation from commensurability:
\begin{align}
\Delta = \frac{j-1}{j}\frac{P_{i+1}}{P_i} - 1.
\end{align}
If the observed young planets represented the pristine outcomes of convergent disk migration, then we expect $\Delta \sim +10^{-3}$ for typical disk aspect ratios \citep{choksi_chiang_2020}.
Instead, observed systems pile up at a larger mean $\Delta \sim +10^{-2}$. At the same time, the young systems V1298 Tau, AU Mic,\footnote{AU~Mic~b and c have a period ratio 1\% below 9:4. TTVs point to a third planet, AU~Mic~d, orbiting between them and completing a chain of 3:2 resonances. Its period is not precisely known because of degeneracies in interpreting the TTV signal \citep{wittrock_etal_2023}. But since b and c together lie just short of (3:2)$^2$, at least one adjacent pair, and possibly both, must lie narrow of exact 3:2.} and TOI-1136 (ages 20, 20, and 700 Myr, respectively) contain pairs with $\Delta < 0$. Both features can be explained by scatterings and collisions with a local population of small bodies. As pointed out by \cite{wu_etal_2024} and \cite{hadden_wu_2026}, scattering shifts the mean $\Delta$ to more positive values. The $\Delta < 0$ outliers may be generated as these small bodies finally accrete and occasionally nudge a pair together. A collision with one super-Mercury is enough to displace a pair of super-Earths to the narrow side of resonance (Fig.~\ref{fig:fid_periodratio}). Accretion of many smaller bodies \citep{hadden_wu_2026} cannot do this, because the effects of their impacts cancel out. 

A related property is the free eccentricity of planets near resonance. Dissipative capture into resonance should zero out free eccentricity \citep[][]{choksi_chiang_2023, goldberg_batygin_2023}. But from analysis of transit timing variations (TTVs), \cite{livingston_etal_2026} report that the young V1298 Tau planets possess free eccentricities of order a percent. Mature planets surviving near resonance have typical free eccentricities of a few percent \citep{lithwick_wu_2012, wu_lithwick_2013, hadden_lithwick_2017}. These may have been excited by impacts from small bodies (Fig.~\ref{fig:ecc_msmall}), leaving resonant systems perched on the threshold for stage-two instability.

\newrev{A new observational trend we uncovered is that the proportion of systems with just one transiting planet increases with age (Fig.~\ref{fig:age_multiplicity}). We suggest this trend is caused by the disruption of resonant chains. After a chain undergoes dynamical instability, it leaves behind a dynamically hot system that would be unlikely to host multiple transiting planets (Figs.~\ref{fig:nres} and \ref{fig:fmulti_evolution}). The present-day culmination of this process might then be the ``Kepler dichotomy,'' wherein an excess of singly transiting systems hints at a population with high mutual inclinations and another nearly coplanar population \citep[e.g.][]{lissauer_etal_2011}. Relatedly, this picture would explain why eccentricities inferred from transit durations (\citealt{xie_etal_2016, mills_etal_2019b}), which mostly probe the nonresonant population, are higher than those inferred from TTVs \citep{lithwick_etal_2012, wu_lithwick_2013, hadden_lithwick_2017}, which probe the resonant population. Future studies should perform mock observations \citep[e.g.][]{hansen_murray_2013, he_etal_2019} to try to quantitatively reproduce the observed multiplicity statistics as a function of age.}


\newrev{In our simulations, unstable chains disrupt completely rather than breaking off just a couple of their links (see also \citealt{goldberg_etal_2025} and \citealt{hadden_wu_2026}, who report similar behavior). Figures \ref{fig:resfrac} and \ref{fig:periodratio_multiplicity} offer some observational support for such all-or-nothing disruption. Fig.~\ref{fig:resfrac} plots the fraction of planets that are near resonance in a system as a function of its multiplicity. In high-multiplicity systems, nearly all of the planets are found close to resonance. Such systems drive the excess near 3:2 seen in the histogram of period ratios plotted in Fig.~\ref{fig:periodratio_multiplicity}. But as multiplicity decreases to $N \leq 5$, we find a sharp drop in the occupation of resonances. In these low-multiplicity systems, period ratios distribute more randomly and resonance occupation seems consistent with chance. To show that, we use a simple Monte Carlo model in which we draw $N$ planets randomly from a log-uniform distribution between 10 days and 365 days. The maximum orbital period of 365 days is chosen to match that of the most distant planets \textit{Kepler} could have detected. The log-uniform distribution is modeled after observations (fig.~1 of \citealt{elee_chiang_2017}) and is also obtained in simulations after periods of dynamical instability \citep{marimbu_lee_2024}. To ensure the modeled system is not manifestly unstable, we also discard draws in which a pair of planets lies within 10 mutual Hill radii of each other. From the remaining mock systems, we calculate how often pairs coincidentally position near resonance. Fig.~\ref{fig:resfrac} shows that this ``randomized'' model fits the data at low multiplicities. We interpret this success, together with the resonant excess at high multiplicities, to suggest that systems formed in long resonant chains that later suffered wholesale instabilities that scrambled their orbital periods and reduced their multiplicities.
}

At present, the longest chain known at ages $<100$ Myr has just four planets (fig.~6 of \citealt{dai_etal_2024}). But given the short baseline of \textit{TESS} and the challenge of identifying transits against the activity of young stars, it seems likely that observations of young planetary systems are far from complete. We predict that more exhaustive searches will find additional links to these young chains, making them more similar in length to surviving resonant chains like TRAPPIST-1 or TOI-1136 found around older stars. \newrev{Future work could also try to pin down the distribution of initial chain lengths from the theory side by mock observing simulated systems, constrained against Figs.~\ref{fig:resfrac}-\ref{fig:periodratio_multiplicity}  (see also \citealt{li_etal_2025}, who experiment with combinations of four and six-planet chains).}

\subsection{Origin of small bodies}
\red{We found that resonant super-Earths could be destabilized on $\sim$100 Myr timescales by the accretion of a handful of Mercury-sized bodies totaling a few percent of the planetary system mass. We propose an idea for how these small bodies formed. The rocky cores of planets are thought to grow through a series of successive mergers of smaller protocores in a depleting gas disk \citep[e.g.][]{kominami_ida_2002, elee_chiang_2016}. The giant impacts triggered by the disruption of resonant chains may be just the last in a long series of mergers. The small bodies may grow out of the collisional debris sprayed out into interplanetary space by these earlier epochs of impacts. Close to the star, impacts occur at relative velocities $ev_{\rm K} \sim 10$--30 km/s, just exceeding planetary surface escape speeds of $\sim$10 km/s. In this marginally gravitationally focused regime, the liberated mass is a few percent of the total colliding mass \citep[e.g.][]{leinhardt_stewart_2012, emsenhuber_etal_2024}, comparable to the total mass $M_{\rm small} \sim 0.05M_{\rm big}$ in our modeled systems. And, as we sketch below, this debris could coagulate into bodies of about the right size to break the chains.}

\red{
Initially launched onto eccentric and inclined orbits similar to those of the parent protocores, debris bodies may encounter each other at relative velocities exceeding km/s. Collisions at such high speeds shatter solids and drive a collisional cascade (e.g.~\citealt{wyatt_2008}). Whether all the mass that cascades to smaller sizes is blown away by stellar radiation pressure and/or winds is unclear (more on this issue at the end of this paper). We assume that it is not --- that an order-unity fraction of the solid debris survives in an optically thick disk in which small bodies frequently collide and have their relative velocities damped by inelastic collisions or drag against residual disk gas. In our scenario, the debris need only persist long enough and collide at speeds slow enough to reaccumulate. Larger, self-gravitating debris chunks will sweep up the smaller bodies. The timescale for a body of radius $s$ to double its mass from agglomerative collisions is \citep[e.g.][]{goldreich_etal_2004}:
\begin{align}
t_{\rm col} &\sim \frac{\rho_{\rm int}s}{\Sigma_{\rm solid}}P 
\label{eqn:tcol}
\end{align}
where $\Sigma_{\rm solid} = M_{\rm small}/\pi a^2$ is the surface density of the debris, of total mass $M_{\rm small}$, spread across the orbits of the planets, and $\rho_{\rm int}$ is the bulk density of solid bodies. Growth is limited by the time it takes the super-Earths to accrete the debris, $t_{\rm acc} \sim (a/r_{\rm big})^2P$. Setting the two timescales equal gives the mass to which debris bodies can grow:
\begin{align}
m_{\rm grow} &\sim \left(\frac{a}{r_{\rm big}} \right)^6 \frac{\Sigma_{\rm solid}^3}{\rho_{\rm int}^2} \nonumber \\ 
&\sim 2\Mmerc \left(\frac{M_{\rm small}/M_{\rm big}}{0.05}\right)^3\left(\frac{\rho_{\rm int}}{1.5\,\rm g/cm^3}\right)^{-2}\left(\frac{r_{\rm big}}{2\,\Rearth}\right)^{-6}.
\label{eqn:mgrow}
\end{align}
Thus our fiducial debris disk can 
breed super-Mercuries, subject to input parameters. We compare $m_{\rm grow}$ with
\begin{align}
m_{\rm excite} 
&\sim 3\Mmerc \left(\frac{e}{0.02}\right)^2\left(\frac{m_{\rm big}}{4\,\Mearth}\right)\left(\frac{M_{\rm small}/M_{\rm big}}{0.05}\right)^{-1},
\label{eqn:mrattle}
\end{align}
which is the mass the small bodies should individually have if they are to excite eccentricity $e$ in the big bodies. We obtain this expression by inverting the fit to our numerical results in equation \ref{eqn:randomwalk1} (fixing $e_{\rm small} = 0.3$). To induce instability on timescales of order 100 Myr, we found eccentricities $e \approx 2-5\%$ were needed. Figure \ref{fig:grow} compares $m_{\rm grow}$ and $m_{\rm excite}$
for a few values of $e$ in this range. For this scenario to work, the debris mass is constrained to a fraction 5--9\% of the planetary system mass. Encouragingly, this range is compatible with the debris mass fractions released from marginally gravitationally focussed impacts, 
as mentioned at the beginning of this subsection 
(see fig.~9 of \citealt{emsenhuber_etal_2024}).} 

\red{An unresolved issue in this picture is how the debris dynamically cools enough to re-agglomerate. In principle, the debris could grind down to micron sizes, and if it were to remain optically thin to stellar radiation, could be blown away by radiation pressure. But the debris is likely to be optically thick; the vertical optical depth $\Sigma_{\rm solid}/(\rho_{\rm int} s) \gg 1$ for $s \lesssim 1$ m, with the radial optical depth still larger by the ratio of the disk's radial width to its vertical height, and most of the debris may be in the form of mm-cm sized droplets condensed from vapor released from the originating giant impact (e.g.~Ghosh et al., submitted). Collisions are frequent in optically thick disks, and those between bodies of comparable mass can damp relative velocities \citep[for a discussion, see][]{jankovic_etal_2024}. Another way to reduce velocity dispersions is by aerodynamic drag exerted by residual protoplanetary disk gas. The ambient gas density cannot be too large, though, lest gas dynamical friction undo the eccentricity excitation caused by accretion of the small bodies.\footnote{Gas dynamical friction is especially problematic for the origin scenario proposed by \cite{hadden_wu_2026}. They suggest that the small bodies represent planetary embryos that failed to grow into super-Earths. But how does the initial setup of embryos and planets emerge from the gas disk? Dynamical friction might shield the embryos from accretion while the disk is gas-rich. But as the gas gradually disperses, at some point the circularization time $t_{\rm e}$ would exceed the accretion time $t_{\rm acc} \sim 10^4$ yr. The embryos would then 
be accreted by planets 
in the presence of some gas. Since dynamical friction damping times scale inversely with body mass, residual gas would undo the excitation of the big bodies in $10^4\,\mathrm{yr} \times m_{\rm small}/m_{\rm big} \sim 300$ yr.}  We defer a more quantitative treatment of these evolution pathways to future work. If the problem of velocity damping can be overcome, it may be that super-Earth cores grow through 
repeated cycles of resonance capture and giant impacts, ending with the dispersal of the gas disk. 
}

\section*{Acknowledgements}
We thank Sarah Blunt, Fei Dai, Tuhin Ghosh, Sam Hadden, Christian Hellum Bye, Ryan LoRusso, John Livingston, Ruth Murray-Clay, Masahiro Ogihara, Erik Petigura, and Cristobal Petrovich for helpful discussions. \newrev{Dan Fabrycky provided a careful referee report that improved the quality of the paper and led us to revise our interpretation of figures 10 and 11.} NC and RL were supported by Heising-Simons 51 Pegasi b fellowships. YL acknowledges NASA grant 80NSSC23K1262. EC is supported by the Simons Investigator program, NSF AST grant 2205500, and the Miller Institute for Basic Research in Science, University of California Berkeley. This research used the Savio high-performance computing cluster provided by the Berkeley Research Computing program at the University of California, Berkeley; the Quest high-performance computing facility at Northwestern which is jointly supported by the Office of the Provost, the Office for Research, and Northwestern University Information Technology; and the Resnick High-Performance Computing Center, supported by the Resnick Sustainability Institute at the California Institute of Technology. This research has made use of the NASA Exoplanet Archive, which is operated by the California Institute of Technology, under contract with the National Aeronautics and Space Administration under the Exoplanet Exploration Program.

\section*{Data availability}
Data and codes are available upon request of the authors.

\bibliographystyle{mnras}
\bibliography{planets_nick} 

@article{wyatt_2008,
	adsurl = {https://ui.adsabs.harvard.edu/abs/2008ARA&A..46..339W},
	author = {{Wyatt}, M.~C.},
	doi = {10.1146/annurev.astro.45.051806.110525},
	journal = {\araa},
	month = sep,
	pages = {339-383},
	title = {{Evolution of debris disks.}},
	volume = {46},
	year = 2008}

@article{dainese_albrecht_2025,
	adsurl = {https://ui.adsabs.harvard.edu/abs/2025A&A...695A.253D},
	archiveprefix = {arXiv},
	author = {{Dainese}, Silke and {Albrecht}, Simon H.},
	doi = {10.1051/0004-6361/202452904},
	eid = {A253},
	eprint = {2503.02451},
	journal = {\aap},
	month = mar,
	pages = {A253},
	primaryclass = {astro-ph.EP},
	title = {{No robust statistical evidence for a population of water worlds in a 2025 sample of planets orbiting M stars}},
	volume = {695},
	year = 2025}

@article{goldberg_batygin_2023,
	adsurl = {https://ui.adsabs.harvard.edu/abs/2023ApJ...948...12G},
	archiveprefix = {arXiv},
	author = {{Goldberg}, Max and {Batygin}, Konstantin},
	doi = {10.3847/1538-4357/acc9ae},
	eid = {12},
	eprint = {2211.16725},
	journal = {\apj},
	month = may,
	number = {1},
	pages = {12},
	primaryclass = {astro-ph.EP},
	title = {{Dynamics and Origins of the Near-resonant Kepler Planets}},
	volume = {948},
	year = 2023}

@article{emsenhuber_etal_2024,
	adsurl = {https://ui.adsabs.harvard.edu/abs/2024PSJ.....5...59E},
	archiveprefix = {arXiv},
	author = {{Emsenhuber}, Alexandre and {Asphaug}, Erik and {Cambioni}, Saverio and {Gabriel}, Travis S.~J. and {Schwartz}, Stephen R. and {Melikyan}, Robert E. and {Denton}, C. Adeene},
	doi = {10.3847/PSJ/ad2178},
	eid = {59},
	eprint = {2401.17356},
	journal = {\psj},
	month = mar,
	number = {3},
	pages = {59},
	primaryclass = {astro-ph.EP},
	title = {{A New Database of Giant Impacts over a Wide Range of Masses and with Material Strength: A First Analysis of Outcomes}},
	volume = {5},
	year = 2024}

@article{ogihara_etal_2026,
	adsurl = {https://ui.adsabs.harvard.edu/abs/2026ApJ...996...91O},
	archiveprefix = {arXiv},
	author = {{Ogihara}, Masahiro and {Kunitomo}, Masanobu},
	doi = {10.3847/1538-4357/ae1d78},
	eid = {91},
	eprint = {2511.11328},
	journal = {\apj},
	month = jan,
	number = {1},
	pages = {91},
	primaryclass = {astro-ph.EP},
	title = {{Formation and Disruption of Resonant Chains of Super-Earths: Secular Perturbations from Outer Eccentric Embryos}},
	volume = {996},
	year = 2026}

@article{goldberg_etal_2022,
	adsurl = {https://ui.adsabs.harvard.edu/abs/2022AJ....163..201G},
	archiveprefix = {arXiv},
	author = {{Goldberg}, Max and {Batygin}, Konstantin},
	doi = {10.3847/1538-3881/ac5961},
	eid = {201},
	eprint = {2203.00801},
	journal = {\aj},
	month = may,
	number = {5},
	pages = {201},
	primaryclass = {astro-ph.EP},
	title = {{Architectures of Compact Super-Earth Systems Shaped by Instabilities}},
	volume = {163},
	year = 2022}

@article{goldberg_etal_2025,
	adsurl = {https://ui.adsabs.harvard.edu/abs/2025arXiv251111329G},
	archiveprefix = {arXiv},
	author = {{Goldberg}, Max and {Petit}, Antoine C.},
	doi = {10.48550/arXiv.2511.11329},
	eid = {arXiv:2511.11329},
	eprint = {2511.11329},
	journal = {arXiv e-prints},
	month = nov,
	pages = {arXiv:2511.11329},
	primaryclass = {astro-ph.EP},
	title = {{Close-in compact super-Earth systems emerging from resonant chains: slow destabilization by unseen remnants of formation}},
	year = 2025}

@article{li_etal_2025b,
	adsurl = {https://ui.adsabs.harvard.edu/abs/2026ApJ...998L...5L},
	archiveprefix = {arXiv},
	author = {{Li}, Jiaru and {O'Connor}, Christopher E. and {Rasio}, Frederic A.},
	doi = {10.3847/2041-8213/ae3a9a},
	eid = {L5},
	eprint = {2510.18955},
	journal = {\apjl},
	month = feb,
	number = {1},
	pages = {L5},
	primaryclass = {astro-ph.EP},
	title = {{Intruder Alert: Breaking Resonant Chains with Planetesimal Flybys}},
	volume = {998},
	year = 2026}

@article{goldreich_etal_2004,
	adsurl = {https://ui.adsabs.harvard.edu/abs/2004ARA&A..42..549G},
	archiveprefix = {arXiv},
	author = {{Goldreich}, Peter and {Lithwick}, Yoram and {Sari}, Re'em},
	doi = {10.1146/annurev.astro.42.053102.134004},
	eprint = {astro-ph/0405215},
	journal = {\araa},
	month = sep,
	number = {1},
	pages = {549-601},
	primaryclass = {astro-ph},
	title = {{Planet Formation by Coagulation: A Focus on Uranus and Neptune}},
	volume = {42},
	year = 2004}

@article{leleu_etal_2024,
	adsurl = {https://ui.adsabs.harvard.edu/abs/2024A&A...687L...1L},
	archiveprefix = {arXiv},
	author = {{Leleu}, Adrien and {Delisle}, Jean-Baptiste and {Burn}, Remo and {Izidoro}, Andr{\'e} and {Udry}, St{\'e}phane and {Dumusque}, Xavier and {Lovis}, Christophe and {Millholland}, Sarah and {Parc}, L{\'e}na and {Bouchy}, Fran{\c{c}}ois and {Bourrier}, Vincent and {Alibert}, Yann and {Faria}, Jo{\~a}o and {Mordasini}, Christoph and {S{\'e}gransan}, Damien},
	doi = {10.1051/0004-6361/202450587},
	eid = {L1},
	eprint = {2406.18991},
	journal = {\aap},
	month = jul,
	pages = {L1},
	primaryclass = {astro-ph.EP},
	title = {{Resonant sub-Neptunes are puffier}},
	volume = {687},
	year = 2024}

@article{rogers_etal_2023,
	adsurl = {https://ui.adsabs.harvard.edu/abs/2023ApJ...947L..19R},
	archiveprefix = {arXiv},
	author = {{Rogers}, James G. and {Schlichting}, Hilke E. and {Owen}, James E.},
	doi = {10.3847/2041-8213/acc86f},
	eid = {L19},
	eprint = {2301.04321},
	journal = {\apjl},
	month = apr,
	number = {1},
	pages = {L19},
	primaryclass = {astro-ph.EP},
	title = {{Conclusive Evidence for a Population of Water Worlds around M Dwarfs Remains Elusive}},
	volume = {947},
	year = 2023}

@article{hamer_schlaufman_2024,
	adsurl = {https://ui.adsabs.harvard.edu/abs/2024AJ....167...55H},
	archiveprefix = {arXiv},
	author = {{Hamer}, Jacob H. and {Schlaufman}, Kevin C.},
	doi = {10.3847/1538-3881/ad110e},
	eid = {55},
	eprint = {2312.02260},
	journal = {\aj},
	month = feb,
	number = {2},
	pages = {55},
	primaryclass = {astro-ph.EP},
	title = {{Kepler-discovered Multiple-planet Systems near Period Ratios Suggestive of Mean-motion Resonances Are Young}},
	volume = {167},
	year = 2024}

@article{hadden_wu_2026,
	adsurl = {https://ui.adsabs.harvard.edu/abs/2026arXiv260221349H},
	archiveprefix = {arXiv},
	author = {{Hadden}, Sam and {Wu}, Yanqin},
	eid = {arXiv:2602.21349},
	eprint = {2602.21349},
	journal = {arXiv e-prints},
	month = feb,
	pages = {arXiv:2602.21349},
	primaryclass = {astro-ph.EP},
	title = {{Rattle-and-Break: the Impact of Planetesimal Scattering on Super-Earth Resonant Chains}},
	year = 2026}

@article{dai_etal_2024,
	adsurl = {https://ui.adsabs.harvard.edu/abs/2024AJ....168..239D},
	archiveprefix = {arXiv},
	author = {{Dai}, Fei and {Goldberg}, Max and {Batygin}, Konstantin and {van Saders}, Jennifer and {Chiang}, Eugene and {Choksi}, Nick and {Li}, Rixin and {Petigura}, Erik A. and {Gilbert}, Gregory J. and {Millholland}, Sarah C. and {Dai}, Yuan-Zhe and {Bouma}, Luke and {Weiss}, Lauren M. and {Winn}, Joshua N.},
	doi = {10.3847/1538-3881/ad83a6},
	eid = {239},
	eprint = {2406.06885},
	journal = {\aj},
	month = dec,
	number = {6},
	pages = {239},
	primaryclass = {astro-ph.EP},
	title = {{The Prevalence of Resonance Among Young, Close-in Planets}},
	volume = {168},
	year = 2024}

@article{jankovic_etal_2024,
	adsurl = {https://ui.adsabs.harvard.edu/abs/2024A&A...691A.302J},
	archiveprefix = {arXiv},
	author = {{Jankovic}, Marija R. and {Wyatt}, Mark C. and {L{\"o}hne}, Torsten},
	doi = {10.1051/0004-6361/202451080},
	eid = {A302},
	eprint = {2411.13991},
	journal = {\aap},
	month = nov,
	pages = {A302},
	primaryclass = {astro-ph.EP},
	title = {{Collisional damping in debris discs: Only significant if collision velocities are low}},
	volume = {691},
	year = 2024}

@ARTICLE{marimbu_lee_2024,
       author = {{Marimbu}, Kevin and {Lee}, Eve J.},
        title = "{On the Dynamical Erasure of Initial Conditions in Multi-planetary Systems}",
      journal = {Research Notes of the American Astronomical Society},
         year = 2024,
        month = aug,
       volume = {8},
       number = {8},
          eid = {208},
        pages = {208},
          doi = {10.3847/2515-5172/ad7380},
archivePrefix = {arXiv},
       eprint = {2410.14782},
 primaryClass = {astro-ph.EP},
       adsurl = {https://ui.adsabs.harvard.edu/abs/2024RNAAS...8..208M}
}

@ARTICLE{he_etal_2019,
       author = {{He}, Matthias Y. and {Ford}, Eric B. and {Ragozzine}, Darin},
        title = "{Architectures of exoplanetary systems - I. A clustered forward model for exoplanetary systems around Kepler's FGK stars}",
      journal = {\mnras},
         year = 2019,
        month = dec,
       volume = {490},
       number = {4},
        pages = {4575-4605},
          doi = {10.1093/mnras/stz2869},
archivePrefix = {arXiv},
       eprint = {1907.07773},
 primaryClass = {astro-ph.EP},
       adsurl = {https://ui.adsabs.harvard.edu/abs/2019MNRAS.490.4575H}
}

@article{leinhardt_stewart_2012,
	adsurl = {https://ui.adsabs.harvard.edu/abs/2012ApJ...745...79L},
	archiveprefix = {arXiv},
	author = {{Leinhardt}, Zo{\"e} M. and {Stewart}, Sarah T.},
	doi = {10.1088/0004-637X/745/1/79},
	eid = {79},
	eprint = {1106.6084},
	journal = {\apj},
	month = jan,
	number = {1},
	pages = {79},
	primaryclass = {astro-ph.EP},
	title = {{Collisions between Gravity-dominated Bodies. I. Outcome Regimes and Scaling Laws}},
	volume = {745},
	year = 2012}

@article{holman_wisdom_1993,
	adsurl = {https://ui.adsabs.harvard.edu/abs/1993AJ....105.1987H},
	author = {{Holman}, M.~J. and {Wisdom}, J.},
	doi = {10.1086/116574},
	journal = {\aj},
	month = may,
	pages = {1987},
	title = {{Dynamical Stability in the Outer Solar System and the Delivery of Short Period Comets}},
	volume = {105},
	year = 1993}

@article{lopez-murillo_etal_2026,
	adsurl = {https://ui.adsabs.harvard.edu/abs/2026AJ....171...63L},
	archiveprefix = {arXiv},
	author = {{Lopez Murillo}, Ana Isabel and {Mann}, Andrew W. and {Barber}, Madyson G. and {Vanderburg}, Andrew and {Thao}, Pa Chia and {Boyle}, Andrew W.},
	doi = {10.3847/1538-3881/ae231a},
	eid = {63},
	eprint = {2512.06035},
	journal = {\aj},
	month = feb,
	number = {2},
	pages = {63},
	primaryclass = {astro-ph.EP},
	title = {{Searching for Transit Timing Variations in Young Transiting Systems}},
	volume = {171},
	year = 2026}

@article{lorusso_etal_2026,
	adsurl = {https://ui.adsabs.harvard.edu/abs/2026arXiv260408383L},
	archiveprefix = {arXiv},
	author = {{LoRusso}, Ryan and {Petrovich}, Cristobal and {Gautham Bhaskar}, Hareesh},
	doi = {10.48550/arXiv.2604.08383},
	eid = {arXiv:2604.08383},
	eprint = {2604.08383},
	journal = {arXiv e-prints},
	month = apr,
	pages = {arXiv:2604.08383},
	primaryclass = {astro-ph.EP},
	title = {{Planetesimal-Driven Instabilities in Resonant Chains of Cold Neptunes and Their Dynamical Outcomes}},
	year = 2026}

@article{nagpal_etal_2024,
	adsurl = {https://ui.adsabs.harvard.edu/abs/2024ApJ...969..133N},
	archiveprefix = {arXiv},
	author = {{Nagpal}, Vighnesh and {Goldberg}, Max and {Batygin}, Konstantin},
	doi = {10.3847/1538-4357/ad3046},
	eid = {133},
	eprint = {2403.02412},
	journal = {\apj},
	month = jul,
	number = {2},
	pages = {133},
	primaryclass = {astro-ph.EP},
	title = {{Breaking Giant Chains: Early-stage Instabilities in Long-period Giant Planet Systems}},
	volume = {969},
	year = 2024}

@article{livingston_etal_2026,
	adsurl = {https://ui.adsabs.harvard.edu/abs/2026Natur.649..310L},
	archiveprefix = {arXiv},
	author = {{Livingston}, John H. and {Petigura}, Erik A. and {David}, Trevor J. and {Masuda}, Kento and {Owen}, James and {Nesvorn{\'y}}, David and {Batygin}, Konstantin and {de Leon}, Jerome and {Mori}, Mayuko and {Ikuta}, Kai and {Fukui}, Akihiko and {Watanabe}, Noriharu and {Orell Miquel}, Jaume and {Murgas}, Felipe and {Parviainen}, Hannu and {Korth}, Judith and {Libotte}, Florence and {Abreu Garc{\'\i}a}, N{\'e}stor and {Gallardo}, Pedro Pablo Meni and {Narita}, Norio and {Pall{\'e}}, Enric and {Tamura}, Motohide and {Yonehara}, Atsunori and {Ridden-Harper}, Andrew and {Bieryla}, Allyson and {Trani}, Alessandro A. and {Mamajek}, Eric E. and {Ciardi}, David R. and {Gorjian}, Varoujan and {Hillenbrand}, Lynne A. and {Rebull}, Luisa M. and {Newton}, Elisabeth R. and {Mann}, Andrew W. and {Vanderburg}, Andrew and {Stef{\'a}nsson}, Gu{\dh}mundur and {Mahadevan}, Suvrath and {Ca{\~n}as}, Caleb and {Ninan}, Joe and {Higuera}, Jesus and {Todorov}, Kamen and {D{\'e}sert}, Jean-Michel and {Pino}, Lorenzo},
	doi = {10.1038/s41586-025-09840-z},
	eprint = {2601.10598},
	journal = {\nat},
	month = jan,
	number = {8096},
	pages = {310-314},
	primaryclass = {astro-ph.EP},
	title = {{A young progenitor for the most common planetary systems in the Galaxy}},
	volume = {649},
	year = 2026}

@article{wu_etal_2024,
	adsurl = {https://ui.adsabs.harvard.edu/abs/2024ApJ...971....5W},
	archiveprefix = {arXiv},
	author = {{Wu}, Yanqin and {Malhotra}, Renu and {Lithwick}, Yoram},
	doi = {10.3847/1538-4357/ad5a09},
	eid = {5},
	eprint = {2405.08893},
	journal = {\apj},
	month = aug,
	number = {1},
	pages = {5},
	primaryclass = {astro-ph.EP},
	title = {{Repelling Planet Pairs by Ping-pong Scattering}},
	volume = {971},
	year = 2024}

@article{rogers_owen_2021,
	adsurl = {https://ui.adsabs.harvard.edu/abs/2021MNRAS.503.1526R},
	archiveprefix = {arXiv},
	author = {{Rogers}, James G. and {Owen}, James E.},
	doi = {10.1093/mnras/stab529},
	eprint = {2007.11006},
	journal = {\mnras},
	month = may,
	number = {1},
	pages = {1526-1542},
	primaryclass = {astro-ph.EP},
	title = {{Unveiling the planet population at birth}},
	volume = {503},
	year = 2021}

@article{wu_lithwick_2013,
	adsurl = {https://ui.adsabs.harvard.edu/abs/2013ApJ...772...74W},
	archiveprefix = {arXiv},
	author = {{Wu}, Yanqin and {Lithwick}, Yoram},
	doi = {10.1088/0004-637X/772/1/74},
	eid = {74},
	eprint = {1210.7810},
	journal = {\apj},
	month = jul,
	number = {1},
	pages = {74},
	primaryclass = {astro-ph.EP},
	title = {{Density and Eccentricity of Kepler Planets}},
	volume = {772},
	year = 2013}

@article{choksi_chiang_2023,
	adsurl = {https://ui.adsabs.harvard.edu/abs/2023MNRAS.522.1914C},
	archiveprefix = {arXiv},
	author = {{Choksi}, Nick and {Chiang}, Eugene},
	doi = {10.1093/mnras/stad835},
	eprint = {2211.15701},
	journal = {\mnras},
	month = jun,
	number = {2},
	pages = {1914-1929},
	primaryclass = {astro-ph.EP},
	title = {{Exciting the transit timing variation phases of resonant sub-Neptunes}},
	volume = {522},
	year = 2023}

@article{hu_etal_2025,
	adsurl = {https://ui.adsabs.harvard.edu/abs/2025ApJ...995..206H},
	archiveprefix = {arXiv},
	author = {{Hu}, Zhecheng and {Dai}, Fei and {Zhu}, Wei and {Wang}, Mu-Tian and {Goldberg}, Max and {Lammers}, Caleb and {Masuda}, Kento},
	doi = {10.3847/1538-4357/ae173c},
	eid = {206},
	eprint = {2510.20185},
	journal = {\apj},
	month = dec,
	number = {2},
	pages = {206},
	primaryclass = {astro-ph.EP},
	title = {{Unexpected Near-Resonant and Metastable States of Young Multiplanet Systems}},
	volume = {995},
	year = 2025}

@article{lithwick_etal_2012,
	adsurl = {https://ui.adsabs.harvard.edu/abs/2012ApJ...761..122L},
	archiveprefix = {arXiv},
	author = {{Lithwick}, Yoram and {Xie}, Jiwei and {Wu}, Yanqin},
	doi = {10.1088/0004-637X/761/2/122},
	eid = {122},
	eprint = {1207.4192},
	journal = {\apj},
	month = dec,
	number = {2},
	pages = {122},
	primaryclass = {astro-ph.EP},
	title = {{Extracting Planet Mass and Eccentricity from TTV Data}},
	volume = {761},
	year = 2012}

@article{rein_etal_2019,
	adsurl = {https://ui.adsabs.harvard.edu/abs/2019MNRAS.485.5490R},
	archiveprefix = {arXiv},
	author = {{Rein}, Hanno and {Hernandez}, David M. and {Tamayo}, Daniel and {Brown}, Garett and {Eckels}, Emily and {Holmes}, Emma and {Lau}, Michelle and {Leblanc}, R{\'e}jean and {Silburt}, Ari},
	doi = {10.1093/mnras/stz769},
	eprint = {1903.04972},
	journal = {\mnras},
	month = jun,
	number = {4},
	pages = {5490-5497},
	primaryclass = {astro-ph.EP},
	title = {{Hybrid symplectic integrators for planetary dynamics}},
	volume = {485},
	year = 2019}

@article{rein_spiegel_2015,
	adsurl = {https://ui.adsabs.harvard.edu/abs/2015MNRAS.446.1424R},
	archiveprefix = {arXiv},
	author = {{Rein}, Hanno and {Spiegel}, David S.},
	doi = {10.1093/mnras/stu2164},
	eprint = {1409.4779},
	journal = {\mnras},
	month = jan,
	number = {2},
	pages = {1424-1437},
	primaryclass = {astro-ph.EP},
	title = {{IAS15: a fast, adaptive, high-order integrator for gravitational dynamics, accurate to machine precision over a billion orbits}},
	volume = {446},
	year = 2015}

@article{borderies_goldreich_1984,
	adsurl = {https://ui.adsabs.harvard.edu/abs/1984CeMec..32..127B},
	author = {{Borderies}, N. and {Goldreich}, P.},
	doi = {10.1007/BF01231120},
	journal = {Celestial Mechanics},
	month = feb,
	number = {2},
	pages = {127-136},
	title = {{A Simple Derivation of Capture Probabilities for the J+1:J and J+2:J Orbit-Orbit Resonance Problems}},
	volume = {32},
	year = 1984}

@article{pu_wu_2015,
	adsurl = {https://ui.adsabs.harvard.edu/abs/2015ApJ...807...44P},
	archiveprefix = {arXiv},
	author = {{Pu}, Bonan and {Wu}, Yanqin},
	doi = {10.1088/0004-637X/807/1/44},
	eid = {44},
	eprint = {1502.05449},
	journal = {\apj},
	month = jul,
	number = {1},
	pages = {44},
	primaryclass = {astro-ph.EP},
	title = {{Spacing of Kepler Planets: Sculpting by Dynamical Instability}},
	volume = {807},
	year = 2015}

@article{hansen_etal_2025,
	adsurl = {https://ui.adsabs.harvard.edu/abs/2025arXiv251001332H},
	archiveprefix = {arXiv},
	author = {{Hansen}, Brad M.~S. and {Yu}, Tze-Yeung and {Nagarajan}, Neel and {Hasegawa}, Yasuhiro},
	doi = {10.48550/arXiv.2510.01332},
	eid = {arXiv:2510.01332},
	eprint = {2510.01332},
	journal = {arXiv e-prints},
	month = oct,
	pages = {arXiv:2510.01332},
	primaryclass = {astro-ph.EP},
	title = {{Dynamical Excitation as a probe of planetary origins}},
	year = 2025}

@article{tamayo_etal_2020,
	adsurl = {https://ui.adsabs.harvard.edu/abs/2020MNRAS.491.2885T},
	archiveprefix = {arXiv},
	author = {{Tamayo}, Daniel and {Rein}, Hanno and {Shi}, Pengshuai and {Hernandez}, David M.},
	doi = {10.1093/mnras/stz2870},
	eprint = {1908.05634},
	journal = {\mnras},
	month = jan,
	number = {2},
	pages = {2885-2901},
	primaryclass = {astro-ph.EP},
	title = {{REBOUNDx: a library for adding conservative and dissipative forces to otherwise symplectic N-body integrations}},
	volume = {491},
	year = 2020}

@article{batygin_2015,
	adsurl = {https://ui.adsabs.harvard.edu/abs/2015MNRAS.451.2589B},
	archiveprefix = {arXiv},
	author = {{Batygin}, Konstantin},
	doi = {10.1093/mnras/stv1063},
	eprint = {1505.01778},
	journal = {\mnras},
	month = aug,
	number = {3},
	pages = {2589-2609},
	primaryclass = {astro-ph.EP},
	title = {{Capture of planets into mean-motion resonances and the origins of extrasolar orbital architectures}},
	volume = {451},
	year = 2015}

@article{mann_etal_2020,
	adsurl = {https://ui.adsabs.harvard.edu/abs/2020AJ....160..179M},
	archiveprefix = {arXiv},
	author = {{Mann}, Andrew W. and {Johnson}, Marshall C. and {Vanderburg}, Andrew and {Kraus}, Adam L. and {Rizzuto}, Aaron C. and {Wood}, Mackenna L. and {Bush}, Jonathan L. and {Rockcliffe}, Keighley and {Newton}, Elisabeth R. and {Latham}, David W. and {Mamajek}, Eric E. and {Zhou}, George and {Quinn}, Samuel N. and {Thao}, Pa Chia and {Benatti}, Serena and {Cosentino}, Rosario and {Desidera}, Silvano and {Harutyunyan}, Avet and {Lovis}, Christophe and {Mortier}, Annelies and {Pepe}, Francesco A. and {Poretti}, Ennio and {Wilson}, Thomas G. and {Kristiansen}, Martti H. and {Gagliano}, Robert and {Jacobs}, Thomas and {LaCourse}, Daryll M. and {Omohundro}, Mark and {Schwengeler}, Hans Martin and {Terentev}, Ivan A. and {Kane}, Stephen R. and {Hill}, Michelle L. and {Rabus}, Markus and {Esquerdo}, Gilbert A. and {Berlind}, Perry and {Collins}, Karen A. and {Murawski}, Gabriel and {Sallam}, Nezar Hazam and {Aitken}, Michael M. and {Massey}, Bob and {Ricker}, George R. and {Vanderspek}, Roland and {Seager}, Sara and {Winn}, Joshua N. and {Jenkins}, Jon M. and {Barclay}, Thomas and {Caldwell}, Douglas A. and {Dragomir}, Diana and {Doty}, John P. and {Glidden}, Ana and {Tenenbaum}, Peter and {Torres}, Guillermo and {Twicken}, Joseph D. and {Villanueva}, Jr., Steven},
	doi = {10.3847/1538-3881/abae64},
	eid = {179},
	eprint = {2005.00047},
	journal = {\aj},
	month = oct,
	number = {4},
	pages = {179},
	primaryclass = {astro-ph.EP},
	title = {{TESS Hunt for Young and Maturing Exoplanets (THYME). III. A Two-planet System in the 400 Myr Ursa Major Group}},
	volume = {160},
	year = 2020}

@article{thao_etal_2024,
	adsurl = {https://ui.adsabs.harvard.edu/abs/2024AJ....168...41T},
	archiveprefix = {arXiv},
	author = {{Thao}, Pa Chia and {Mann}, Andrew W. and {Barber}, Madyson G. and {Kraus}, Adam L. and {Tofflemire}, Benjamin M. and {Bush}, Jonathan L. and {Wood}, Mackenna L. and {Collins}, Karen A. and {Vanderburg}, Andrew and {Quinn}, Samuel N. and {Zhou}, George and {Newton}, Elisabeth R. and {Ziegler}, Carl and {Law}, Nicholas and {Barkaoui}, Khalid and {Pozuelos}, Francisco J. and {Timmermans}, Mathilde and {Gillon}, Micha{\"e}l and {Jehin}, Emmanu{\"e}l and {Schwarz}, Richard P. and {Gan}, Tianjun and {Shporer}, Avi and {Horne}, Keith and {Sefako}, Ramotholo and {Suarez}, Olga and {Mekarnia}, Djamel and {Guillot}, Tristan and {Abe}, Lyu and {Triaud}, Amaury H.~M.~J. and {Radford}, Don J. and {Lopez Murillo}, Ana Isabel and {Ricker}, George R. and {Winn}, Joshua N. and {Jenkins}, Jon M. and {Bouma}, Luke G. and {Fausnaugh}, Michael and {Guerrero}, Natalia M. and {Kunimoto}, Michelle},
	doi = {10.3847/1538-3881/ad4993},
	eid = {41},
	eprint = {2406.05234},
	journal = {\aj},
	month = jul,
	number = {1},
	pages = {41},
	primaryclass = {astro-ph.EP},
	title = {{TESS Hunt for Young and Maturing Exoplanets (THYME). X. A Two-planet System in the 210 Myr MELANGE-5 Association}},
	volume = {168},
	year = 2024}

@article{newton_etal_2021,
	adsurl = {https://ui.adsabs.harvard.edu/abs/2021AJ....161...65N},
	archiveprefix = {arXiv},
	author = {{Newton}, Elisabeth R. and {Mann}, Andrew W. and {Kraus}, Adam L. and {Livingston}, John H. and {Vanderburg}, Andrew and {Curtis}, Jason L. and {Thao}, Pa Chia and {Hawkins}, Keith and {Wood}, Mackenna L. and {Rizzuto}, Aaron C. and {Soubkiou}, Abderahmane and {Tofflemire}, Benjamin M. and {Zhou}, George and {Crossfield}, Ian J.~M. and {Pearce}, Logan A. and {Collins}, Karen A. and {Conti}, Dennis M. and {Tan}, Thiam-Guan and {Villeneuva}, Steven and {Spencer}, Alton and {Dragomir}, Diana and {Quinn}, Samuel N. and {Jensen}, Eric L.~N. and {Collins}, Kevin I. and {Stockdale}, Chris and {Cloutier}, Ryan and {Hellier}, Coel and {Benkhaldoun}, Zouhair and {Ziegler}, Carl and {Brice{\~n}o}, C{\'e}sar and {Law}, Nicholas and {Benneke}, Bj{\"o}rn and {Christiansen}, Jessie L. and {Gorjian}, Varoujan and {Kane}, Stephen R. and {Kreidberg}, Laura and {Morales}, Farisa Y. and {Werner}, Michael W. and {Twicken}, Joseph D. and {Levine}, Alan M. and {Ciardi}, David R. and {Guerrero}, Natalia M. and {Hesse}, Katharine and {Quintana}, Elisa V. and {Shiao}, Bernie and {Smith}, Jeffrey C. and {Torres}, Guillermo and {Ricker}, George R. and {Vanderspek}, Roland and {Seager}, Sara and {Winn}, Joshua N. and {Jenkins}, Jon M. and {Latham}, David W.},
	doi = {10.3847/1538-3881/abccc6},
	eid = {65},
	eprint = {2102.06049},
	journal = {\aj},
	month = feb,
	number = {2},
	pages = {65},
	primaryclass = {astro-ph.EP},
	title = {{TESS Hunt for Young and Maturing Exoplanets (THYME). IV. Three Small Planets Orbiting a 120 Myr Old Star in the Pisces-Eridanus Stream}},
	volume = {161},
	year = 2021}

@article{dattilo_etal_2025,
	adsurl = {https://ui.adsabs.harvard.edu/abs/2025AJ....170..318D},
	archiveprefix = {arXiv},
	author = {{Dattilo}, Anne and {Vanderburg}, Andrew M. and {Barber}, Madyson G. and {Mann}, Andrew W. and {Kerr}, Ronan and {Kraus}, Adam L. and {Livesey}, Joseph R. and {Watkins}, Cristilyn and {Collins}, Karen A. and {Garc{\'\i}a-Mej{\'\i}a}, Juliana and {Tamburo}, Patrick and {Becker}, Juliette and {Mortier}, Annelies and {Wilson}, Thomas and {Scarsdale}, Nicholas and {Gilbert}, Emily A. and {Polanski}, Alex S. and {Howell}, Steve B. and {Crossfield}, Ian and {Bieryla}, Allyson and {Ciardi}, David R. and {Barclay}, Thomas and {Charbonneau}, David and {Latham}, David W. and {Murphy}, Joseph M. Akana and {Newton}, Elisabeth and {Massey}, Bob and {Schwarz}, Richard P. and {Stockdale}, Chris and {Wilkin}, Francis P. and {Zambelli}, Roberto},
	doi = {10.3847/1538-3881/ae0a35},
	eid = {318},
	eprint = {2509.15313},
	journal = {\aj},
	month = dec,
	number = {6},
	pages = {318},
	primaryclass = {astro-ph.EP},
	title = {{THYME. XIII. Two Young Neptunes Orbiting a 75 Myr Star in the Alpha Persei Cluster}},
	volume = {170},
	year = 2025}

@inproceedings{ricker_etal_2014,
	adsurl = {https://ui.adsabs.harvard.edu/abs/2014SPIE.9143E..20R},
	archiveprefix = {arXiv},
	author = {{Ricker}, George R. and {Winn}, Joshua N. and {Vanderspek}, Roland and {Latham}, David W. and {Bakos}, G{\'a}sp{\'a}r. {\'A}. and {Bean}, Jacob L. and {Berta-Thompson}, Zachory K. and {Brown}, Timothy M. and {Buchhave}, Lars and {Butler}, Nathaniel R. and {Butler}, R. Paul and {Chaplin}, William J. and {Charbonneau}, David and {Christensen-Dalsgaard}, J{\o}rgen and {Clampin}, Mark and {Deming}, Drake and {Doty}, John and {De Lee}, Nathan and {Dressing}, Courtney and {Dunham}, E.~W. and {Endl}, Michael and {Fressin}, Francois and {Ge}, Jian and {Henning}, Thomas and {Holman}, Matthew J. and {Howard}, Andrew W. and {Ida}, Shigeru and {Jenkins}, Jon and {Jernigan}, Garrett and {Johnson}, John A. and {Kaltenegger}, Lisa and {Kawai}, Nobuyuki and {Kjeldsen}, Hans and {Laughlin}, Gregory and {Levine}, Alan M. and {Lin}, Douglas and {Lissauer}, Jack J. and {MacQueen}, Phillip and {Marcy}, Geoffrey and {McCullough}, P.~R. and {Morton}, Timothy D. and {Narita}, Norio and {Paegert}, Martin and {Palle}, Enric and {Pepe}, Francesco and {Pepper}, Joshua and {Quirrenbach}, Andreas and {Rinehart}, S.~A. and {Sasselov}, Dimitar and {Sato}, Bun'ei and {Seager}, Sara and {Sozzetti}, Alessandro and {Stassun}, Keivan G. and {Sullivan}, Peter and {Szentgyorgyi}, Andrew and {Torres}, Guillermo and {Udry}, Stephane and {Villasenor}, Joel},
	booktitle = {Space Telescopes and Instrumentation 2014: Optical, Infrared, and Millimeter Wave},
	doi = {10.1117/12.2063489},
	editor = {{Oschmann}, Jr., Jacobus M. and {Clampin}, Mark and {Fazio}, Giovanni G. and {MacEwen}, Howard A.},
	eid = {914320},
	eprint = {1406.0151},
	month = aug,
	pages = {914320},
	primaryclass = {astro-ph.EP},
	series = {Society of Photo-Optical Instrumentation Engineers (SPIE) Conference Series},
	title = {{Transiting Exoplanet Survey Satellite (TESS)}},
	volume = {9143},
	year = 2014}

@article{wittrock_etal_2023,
	adsurl = {https://ui.adsabs.harvard.edu/abs/2023AJ....166..232W},
	archiveprefix = {arXiv},
	author = {{Wittrock}, Justin M. and {Plavchan}, Peter P. and {Cale}, Bryson L. and {Barclay}, Thomas and {Ludwig}, Mathis R. and {Schwarz}, Richard P. and {M{\'e}karnia}, Djamel and {Triaud}, Amaury H.~M.~J. and {Abe}, Lyu and {Suarez}, Olga and {Guillot}, Tristan and {Conti}, Dennis M. and {Collins}, Karen A. and {Waite}, Ian A. and {Kielkopf}, John F. and {Collins}, Kevin I. and {Dreizler}, Stefan and {El Mufti}, Mohammed and {Feliz}, Dax L. and {Gaidos}, Eric and {Geneser}, Claire S. and {Horne}, Keith D. and {Kane}, Stephen R. and {Lowrance}, Patrick J. and {Martioli}, Eder and {Radford}, Don J. and {Reefe}, Michael A. and {Roccatagliata}, Veronica and {Shporer}, Avi and {Stassun}, Keivan G. and {Stockdale}, Christopher and {Tan}, Thiam-Guan and {Tanner}, Angelle M. and {Vega}, Laura D.},
	doi = {10.3847/1538-3881/acfda8},
	eid = {232},
	eprint = {2302.04922},
	journal = {\aj},
	month = dec,
	number = {6},
	pages = {232},
	primaryclass = {astro-ph.EP},
	title = {{Validating AU Microscopii d with Transit Timing Variations}},
	volume = {166},
	year = 2023}

@article{newton_etal_2019,
	adsurl = {https://ui.adsabs.harvard.edu/abs/2019ApJ...880L..17N},
	archiveprefix = {arXiv},
	author = {{Newton}, Elisabeth R. and {Mann}, Andrew W. and {Tofflemire}, Benjamin M. and {Pearce}, Logan and {Rizzuto}, Aaron C. and {Vanderburg}, Andrew and {Martinez}, Raquel A. and {Wang}, Jason J. and {Ruffio}, Jean-Baptiste and {Kraus}, Adam L. and {Johnson}, Marshall C. and {Thao}, Pa Chia and {Wood}, Mackenna L. and {Rampalli}, Rayna and {Nielsen}, Eric L. and {Collins}, Karen A. and {Dragomir}, Diana and {Hellier}, Coel and {Anderson}, D.~R. and {Barclay}, Thomas and {Brown}, Carolyn and {Feiden}, Gregory and {Hart}, Rhodes and {Isopi}, Giovanni and {Kielkopf}, John F. and {Mallia}, Franco and {Nelson}, Peter and {Rodriguez}, Joseph E. and {Stockdale}, Chris and {Waite}, Ian A. and {Wright}, Duncan J. and {Lissauer}, Jack J. and {Ricker}, George R. and {Vanderspek}, Roland and {Latham}, David W. and {Seager}, S. and {Winn}, Joshua N. and {Jenkins}, Jon M. and {Bouma}, Luke G. and {Burke}, Christopher J. and {Davies}, Misty and {Fausnaugh}, Michael and {Li}, Jie and {Morris}, Robert L. and {Mukai}, Koji and {Villase{\~n}or}, Joel and {Villeneuva}, Steven and {De Rosa}, Robert J. and {Macintosh}, Bruce and {Mengel}, Matthew W. and {Okumura}, Jack and {Wittenmyer}, Robert A.},
	doi = {10.3847/2041-8213/ab2988},
	eid = {L17},
	eprint = {1906.10703},
	journal = {\apjl},
	month = jul,
	number = {1},
	pages = {L17},
	primaryclass = {astro-ph.EP},
	title = {{TESS Hunt for Young and Maturing Exoplanets (THYME): A Planet in the 45 Myr Tucana-Horologium Association}},
	volume = {880},
	year = 2019}

@article{yang_etal_2023,
	adsurl = {https://ui.adsabs.harvard.edu/abs/2023AJ....166..243Y},
	archiveprefix = {arXiv},
	author = {{Yang}, Jia-Yi and {Chen}, Di-Chang and {Xie}, Ji-Wei and {Zhou}, Ji-Lin and {Dong}, Subo and {Zhu}, Zi and {Zheng}, Zheng and {Liu}, Chao and {Zong}, Weikai and {Luo}, Ali},
	doi = {10.3847/1538-3881/ad0368},
	eid = {243},
	eprint = {2310.20113},
	journal = {\aj},
	month = dec,
	number = {6},
	pages = {243},
	primaryclass = {astro-ph.EP},
	title = {{Planets Across Space and Time (PAST). IV. The Occurrence and Architecture of Kepler Planetary Systems as a Function of Kinematic Age Revealed by the LAMOST-Gaia-Kepler Sample}},
	volume = {166},
	year = 2023}

@article{rein_tamayo_2015,
	adsurl = {https://ui.adsabs.harvard.edu/abs/2015MNRAS.452..376R},
	archiveprefix = {arXiv},
	author = {{Rein}, Hanno and {Tamayo}, Daniel},
	doi = {10.1093/mnras/stv1257},
	eprint = {1506.01084},
	journal = {\mnras},
	month = sep,
	number = {1},
	pages = {376-388},
	primaryclass = {astro-ph.EP},
	title = {{WHFAST: a fast and unbiased implementation of a symplectic Wisdom-Holman integrator for long-term gravitational simulations}},
	volume = {452},
	year = 2015}

@article{rein_liu_2012,
	adsurl = {https://ui.adsabs.harvard.edu/abs/2012A&A...537A.128R},
	archiveprefix = {arXiv},
	author = {{Rein}, H. and {Liu}, S. -F.},
	doi = {10.1051/0004-6361/201118085},
	eid = {A128},
	eprint = {1110.4876},
	journal = {\aap},
	month = jan,
	pages = {A128},
	primaryclass = {astro-ph.EP},
	title = {{REBOUND: an open-source multi-purpose N-body code for collisional dynamics}},
	volume = {537},
	year = 2012}

@article{mills_etal_2019b,
	adsurl = {https://ui.adsabs.harvard.edu/abs/2019AJ....157..198M},
	archiveprefix = {arXiv},
	author = {{Mills}, Sean M. and {Howard}, Andrew W. and {Petigura}, Erik A. and {Fulton}, Benjamin J. and {Isaacson}, Howard and {Weiss}, Lauren M.},
	doi = {10.3847/1538-3881/ab1009},
	eid = {198},
	eprint = {1905.04625},
	journal = {\aj},
	month = may,
	number = {5},
	pages = {198},
	primaryclass = {astro-ph.EP},
	title = {{The California-Kepler Survey. VIII. Eccentricities of Kepler Planets and Tentative Evidence of a High-metallicity Preference for Small Eccentric Planets}},
	volume = {157},
	year = 2019}

@article{xie_etal_2016,
	adsurl = {https://ui.adsabs.harvard.edu/abs/2016PNAS..11311431X},
	archiveprefix = {arXiv},
	author = {{Xie}, Ji-Wei and {Dong}, Subo and {Zhu}, Zhaohuan and {Huber}, Daniel and {Zheng}, Zheng and {De Cat}, Peter and {Fu}, Jianning and {Liu}, Hui-Gen and {Luo}, Ali and {Wu}, Yue and {Zhang}, Haotong and {Zhang}, Hui and {Zhou}, Ji-Lin and {Cao}, Zihuang and {Hou}, Yonghui and {Wang}, Yuefei and {Zhang}, Yong},
	doi = {10.1073/pnas.1604692113},
	eprint = {1609.08633},
	journal = {Proceedings of the National Academy of Science},
	month = oct,
	number = {41},
	pages = {11431-11435},
	primaryclass = {astro-ph.EP},
	title = {{Exoplanet orbital eccentricities derived from LAMOST-Kepler analysis}},
	volume = {113},
	year = 2016}

@article{goldreich_1965,
	adsurl = {https://ui.adsabs.harvard.edu/abs/1965MNRAS.130..159G},
	author = {{Goldreich}, P.},
	doi = {10.1093/mnras/130.3.159},
	journal = {\mnras},
	month = jan,
	pages = {159},
	title = {{An explanation of the frequent occurrence of commensurable mean motions in the solar system}},
	volume = {130},
	year = 1965}

@article{choksi_chiang_2020,
	adsurl = {https://ui.adsabs.harvard.edu/abs/2020MNRAS.495.4192C},
	archiveprefix = {arXiv},
	author = {{Choksi}, Nick and {Chiang}, Eugene},
	doi = {10.1093/mnras/staa1421},
	eprint = {2003.03388},
	journal = {\mnras},
	month = jul,
	number = {4},
	pages = {4192-4209},
	primaryclass = {astro-ph.EP},
	title = {{Sub-Neptune formation: the view from resonant planets}},
	volume = {495},
	year = 2020}

@article{hadden_lithwick_2017,
	adsurl = {https://ui.adsabs.harvard.edu/abs/2017AJ....154....5H},
	archiveprefix = {arXiv},
	author = {{Hadden}, Sam and {Lithwick}, Yoram},
	doi = {10.3847/1538-3881/aa71ef},
	eid = {5},
	eprint = {1611.03516},
	journal = {\aj},
	month = jul,
	number = {1},
	pages = {5},
	primaryclass = {astro-ph.EP},
	title = {{Kepler Planet Masses and Eccentricities from TTV Analysis}},
	volume = {154},
	year = 2017}

@article{zhu_etal_2018,
	adsurl = {https://ui.adsabs.harvard.edu/abs/2018ApJ...860..101Z},
	archiveprefix = {arXiv},
	author = {{Zhu}, Wei and {Petrovich}, Cristobal and {Wu}, Yanqin and {Dong}, Subo and {Xie}, Jiwei},
	doi = {10.3847/1538-4357/aac6d5},
	eid = {101},
	eprint = {1802.09526},
	journal = {\apj},
	month = {Jun},
	number = {2},
	pages = {101},
	primaryclass = {astro-ph.EP},
	title = {{About 30\% of Sun-like Stars Have Kepler-like Planetary Systems: A Study of Their Intrinsic Architecture}},
	volume = {860},
	year = {2018}}

@article{lithwick_wu_2012,
	adsurl = {https://ui.adsabs.harvard.edu/abs/2012ApJ...756L..11L},
	archiveprefix = {arXiv},
	author = {{Lithwick}, Yoram and {Wu}, Yanqin},
	doi = {10.1088/2041-8205/756/1/L11},
	eid = {L11},
	eprint = {1204.2555},
	journal = {\apjl},
	month = {Sep},
	number = {1},
	pages = {L11},
	primaryclass = {astro-ph.EP},
	title = {{Resonant Repulsion of Kepler Planet Pairs}},
	volume = {756},
	year = {2012}}

@article{fabrycky_etal_2014,
	adsurl = {https://ui.adsabs.harvard.edu/abs/2014ApJ...790..146F},
	archiveprefix = {arXiv},
	author = {{Fabrycky}, Daniel C. and {Lissauer}, Jack J. and {Ragozzine}, Darin and {Rowe}, Jason F. and {Steffen}, Jason H. and {Agol}, Eric and {Barclay}, Thomas and {Batalha}, Natalie and {Borucki}, William and {Ciardi}, David R. and {Ford}, Eric B. and {Gautier}, Thomas N. and {Geary}, John C. and {Holman}, Matthew J. and {Jenkins}, Jon M. and {Li}, Jie and {Morehead}, Robert C. and {Morris}, Robert L. and {Shporer}, Avi and {Smith}, Jeffrey C. and {Still}, Martin and {Van Cleve}, Jeffrey},
	doi = {10.1088/0004-637X/790/2/146},
	eid = {146},
	eprint = {1202.6328},
	journal = {\apj},
	month = {Aug},
	number = {2},
	pages = {146},
	primaryclass = {astro-ph.EP},
	title = {{Architecture of Kepler's Multi-transiting Systems. II. New Investigations with Twice as Many Candidates}},
	volume = {790},
	year = {2014}}

@article{terquem_papaloizou_2019,
	adsurl = {https://ui.adsabs.harvard.edu/abs/2019MNRAS.482..530T},
	archiveprefix = {arXiv},
	author = {{Terquem}, Caroline and {Papaloizou}, John C.~B.},
	doi = {10.1093/mnras/sty2693},
	eprint = {1809.10042},
	journal = {\mnras},
	month = {Jan},
	number = {1},
	pages = {530-549},
	primaryclass = {astro-ph.EP},
	title = {{First-order mean motion resonances in two-planet systems: general analysis and observed systems}},
	volume = {482},
	year = {2019}}

@article{kominami_ida_2002,
	adsurl = {https://ui.adsabs.harvard.edu/abs/2002Icar..157...43K},
	author = {{Kominami}, Junko and {Ida}, Shigeru},
	doi = {10.1006/icar.2001.6811},
	journal = {\icarus},
	month = {May},
	number = {1},
	pages = {43-56},
	title = {{The Effect of Tidal Interaction with a Gas Disk on Formation of Terrestrial Planets}},
	volume = {157},
	year = {2002}}

@article{lissauer_etal_2011,
	adsurl = {https://ui.adsabs.harvard.edu/abs/2011ApJS..197....8L},
	archiveprefix = {arXiv},
	author = {{Lissauer}, Jack J. and {Ragozzine}, Darin and {Fabrycky}, Daniel C. and {Steffen}, Jason H. and {Ford}, Eric B. and {Jenkins}, Jon M. and {Shporer}, Avi and {Holman}, Matthew J. and {Rowe}, Jason F. and {Quintana}, Elisa V. and {Batalha}, Natalie M. and {Borucki}, William J. and {Bryson}, Stephen T. and {Caldwell}, Douglas A. and {Carter}, Joshua A. and {Ciardi}, David and {Dunham}, Edward W. and {Fortney}, Jonathan J. and {Gautier}, Thomas N., III and {Howell}, Steve B. and {Koch}, David G. and {Latham}, David W. and {Marcy}, Geoffrey W. and {Morehead}, Robert C. and {Sasselov}, Dimitar},
	doi = {10.1088/0067-0049/197/1/8},
	eid = {8},
	eprint = {1102.0543},
	journal = {\apjs},
	month = {Nov},
	number = {1},
	pages = {8},
	primaryclass = {astro-ph.EP},
	title = {{Architecture and Dynamics of Kepler's Candidate Multiple Transiting Planet Systems}},
	volume = {197},
	year = {2011}}

@article{goldreich_tremaine_1980,
	adsurl = {https://ui.adsabs.harvard.edu/abs/1980ApJ...241..425G},
	author = {{Goldreich}, P. and {Tremaine}, S.},
	doi = {10.1086/158356},
	journal = {\apj},
	month = oct,
	pages = {425-441},
	title = {{Disk-satellite interactions}},
	volume = 241,
	year = 1980}

@article{elee_chiang_2016,
	adsurl = {https://ui.adsabs.harvard.edu/abs/2016ApJ...817...90L},
	archiveprefix = {arXiv},
	author = {{Lee}, Eve J. and {Chiang}, Eugene},
	doi = {10.3847/0004-637X/817/2/90},
	eid = {90},
	eprint = {1510.08855},
	journal = {\apj},
	month = {Feb},
	number = {2},
	pages = {90},
	primaryclass = {astro-ph.EP},
	title = {{Breeding Super-Earths and Birthing Super-puffs in Transitional Disks}},
	volume = {817},
	year = {2016}}

@article{elee_chiang_2017,
	adsurl = {https://ui.adsabs.harvard.edu/abs/2017ApJ...842...40L},
	archiveprefix = {arXiv},
	author = {{Lee}, Eve J. and {Chiang}, Eugene},
	doi = {10.3847/1538-4357/aa6fb3},
	eid = {40},
	eprint = {1702.08461},
	journal = {\apj},
	month = {Jun},
	number = {1},
	pages = {40},
	primaryclass = {astro-ph.EP},
	title = {{Magnetospheric Truncation, Tidal Inspiral, and the Creation of Short-period and Ultra-short-period Planets}},
	volume = {842},
	year = {2017}}

@article{izidoro_etal_2017,
	adsurl = {https://ui.adsabs.harvard.edu/abs/2017MNRAS.470.1750I},
	archiveprefix = {arXiv},
	author = {{Izidoro}, Andre and {Ogihara}, Masahiro and {Raymond}, Sean N. and {Morbidelli}, Alessandro and {Pierens}, Arnaud and {Bitsch}, Bertram and {Cossou}, Christophe and {Hersant}, Franck},
	doi = {10.1093/mnras/stx1232},
	eprint = {1703.03634},
	journal = {\mnras},
	month = {Sep},
	number = {2},
	pages = {1750-1770},
	primaryclass = {astro-ph.EP},
	title = {{Breaking the chains: hot super-Earth systems from migration and disruption of compact resonant chains}},
	volume = {470},
	year = {2017}}

@article{hansen_murray_2013,
	adsurl = {https://ui.adsabs.harvard.edu/abs/2013ApJ...775...53H},
	archiveprefix = {arXiv},
	author = {{Hansen}, Brad M.~S. and {Murray}, Norm},
	doi = {10.1088/0004-637X/775/1/53},
	eid = {53},
	eprint = {1301.7431},
	journal = {\apj},
	month = {Sep},
	number = {1},
	pages = {53},
	primaryclass = {astro-ph.EP},
	title = {{Testing in Situ Assembly with the Kepler Planet Candidate Sample}},
	volume = {775},
	year = {2013}}

@article{dressing_charbonneau_2015,
	adsurl = {https://ui.adsabs.harvard.edu/abs/2015ApJ...807...45D},
	archiveprefix = {arXiv},
	author = {{Dressing}, Courtney D. and {Charbonneau}, David},
	doi = {10.1088/0004-637X/807/1/45},
	eid = {45},
	eprint = {1501.01623},
	journal = {\apj},
	month = {Jul},
	number = {1},
	pages = {45},
	primaryclass = {astro-ph.EP},
	title = {{The Occurrence of Potentially Habitable Planets Orbiting M Dwarfs Estimated from the Full Kepler Dataset and an Empirical Measurement of the Detection Sensitivity}},
	volume = {807},
	year = {2015}}

@article{fressin_etal_2013,
	adsurl = {https://ui.adsabs.harvard.edu/abs/2013ApJ...766...81F},
	archiveprefix = {arXiv},
	author = {{Fressin}, Fran{\c{c}}ois and {Torres}, Guillermo and {Charbonneau}, David and {Bryson}, Stephen T. and {Christiansen}, Jessie and {Dressing}, Courtney D. and {Jenkins}, Jon M. and {Walkowicz}, Lucianne M. and {Batalha}, Natalie M.},
	doi = {10.1088/0004-637X/766/2/81},
	eid = {81},
	eprint = {1301.0842},
	journal = {\apj},
	month = {Apr},
	number = {2},
	pages = {81},
	primaryclass = {astro-ph.EP},
	title = {{The False Positive Rate of Kepler and the Occurrence of Planets}},
	volume = {766},
	year = {2013}}

@article{petigura_etal_2018,
	adsurl = {https://ui.adsabs.harvard.edu/abs/2018AJ....155...89P},
	archiveprefix = {arXiv},
	author = {{Petigura}, Erik A. and {Marcy}, Geoffrey W. and {Winn}, Joshua N. and {Weiss}, Lauren M. and {Fulton}, Benjamin J. and {Howard}, Andrew W. and {Sinukoff}, Evan and {Isaacson}, Howard and {Morton}, Timothy D. and {Johnson}, John Asher},
	doi = {10.3847/1538-3881/aaa54c},
	eid = {89},
	eprint = {1712.04042},
	journal = {\aj},
	month = {Feb},
	number = {2},
	pages = {89},
	primaryclass = {astro-ph.EP},
	title = {{The California-Kepler Survey. IV. Metal-rich Stars Host a Greater Diversity of Planets}},
	volume = {155},
	year = {2018}}

@article{sandford_etal_2019,
	adsurl = {https://ui.adsabs.harvard.edu/abs/2019MNRAS.489.3162S},
	archiveprefix = {arXiv},
	author = {{Sandford}, Emily and {Kipping}, David and {Collins}, Michael},
	doi = {10.1093/mnras/stz2350},
	eprint = {1907.08148},
	journal = {\mnras},
	month = {Nov},
	number = {3},
	pages = {3162-3173},
	primaryclass = {astro-ph.EP},
	title = {{The multiplicity distribution of Kepler's exoplanets}},
	volume = {489},
	year = {2019}}

@article{izidoro_etal_2019,
	adsurl = {https://ui.adsabs.harvard.edu/abs/2019arXiv190208772I},
	archiveprefix = {arXiv},
	author = {{Izidoro}, Andr{\'e} and {Bitsch}, Bertram and {Raymond}, Sean N. and {Johansen}, Anders and {Morbidelli}, Alessandro and {Lambrechts}, Michiel and {Jacobson}, Seth A.},
	eid = {arXiv:1902.08772},
	eprint = {1902.08772},
	journal = {arXiv e-prints},
	month = {Feb},
	pages = {arXiv:1902.08772},
	primaryclass = {astro-ph.EP},
	title = {{Formation of planetary systems by pebble accretion and migration: Hot super-Earth systems from breaking compact resonant chains}},
	year = {2019}}

@article{li_etal_2025,
	adsurl = {https://ui.adsabs.harvard.edu/abs/2025AJ....169..323L},
	archiveprefix = {arXiv},
	author = {{Li}, Rixin and {Chiang}, Eugene and {Choksi}, Nick and {Dai}, Fei},
	doi = {10.3847/1538-3881/adce0c},
	eid = {323},
	eprint = {2408.10206},
	journal = {\aj},
	month = jun,
	number = {6},
	pages = {323},
	primaryclass = {astro-ph.EP},
	title = {{The Resonant Remains of Broken Chains from Major and Minor Mergers}},
	volume = {169},
	year = 2025}

\appendix
\section{Experiments with damping}
\label{sec:appendix_damping}
For most of the simulations in this paper, we ignored dissipation. Here we try variations on our fiducial model that include eccentricity damping using the \texttt{modify-orbits-forces} routine in \texttt{reboundx} \citep{tamayo_etal_2020}.

\begin{figure} 
\includegraphics[width=\columnwidth]{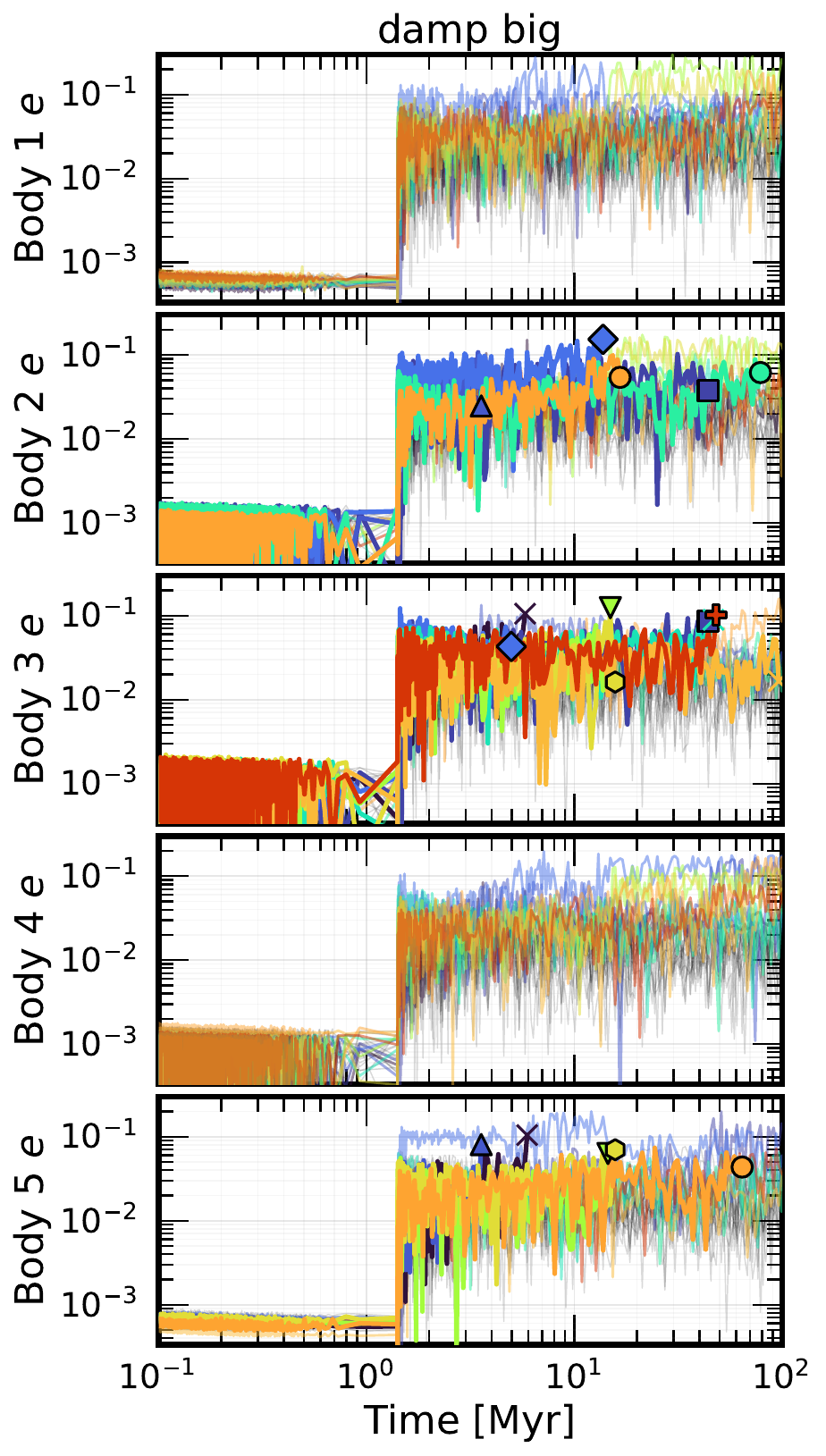}
\vspace{-0.5cm}
\caption{Similar to Fig.~\ref{fig:fidbig}, but now assessing how eccentricity damping on the resonant chain affects stability. In this experiment we start by damping only the big bodies (see text for details). Their eccentricities equilibrate around resonantly forced values of $\sim 10^{-3}$. Around $t=2$ Myr we remove the damping and lay down the small bodies. They excite free eccentricities in the big bodies, which still suffer long-term dynamical instability ending in collisions (marked by symbols). }
\label{fig:damp_big}
\end{figure}

In a first set of experiments, we only damp the big bodies. Our goal is to decide whether they are more stable against impacts from small bodies after having been damped. We integrate in two phases. We start by laying down our fiducial chain and damping their eccentricities on a timescale $t_{e,0} = 10^2$ yr. After 100 $t_{e,0}$, we exponentially decay damping on a timescale $t_{\rm decay} = 10^5\,\rm $ yr. Damping shuts off entirely after 10$t_{\rm decay}$. Then, we lay down the small bodies and continue integrating without dissipation. Figure \ref{fig:damp_big} shows that long-term dynamical instabilities set in despite the applied damping.

\begin{figure} 
\includegraphics[width=\columnwidth]{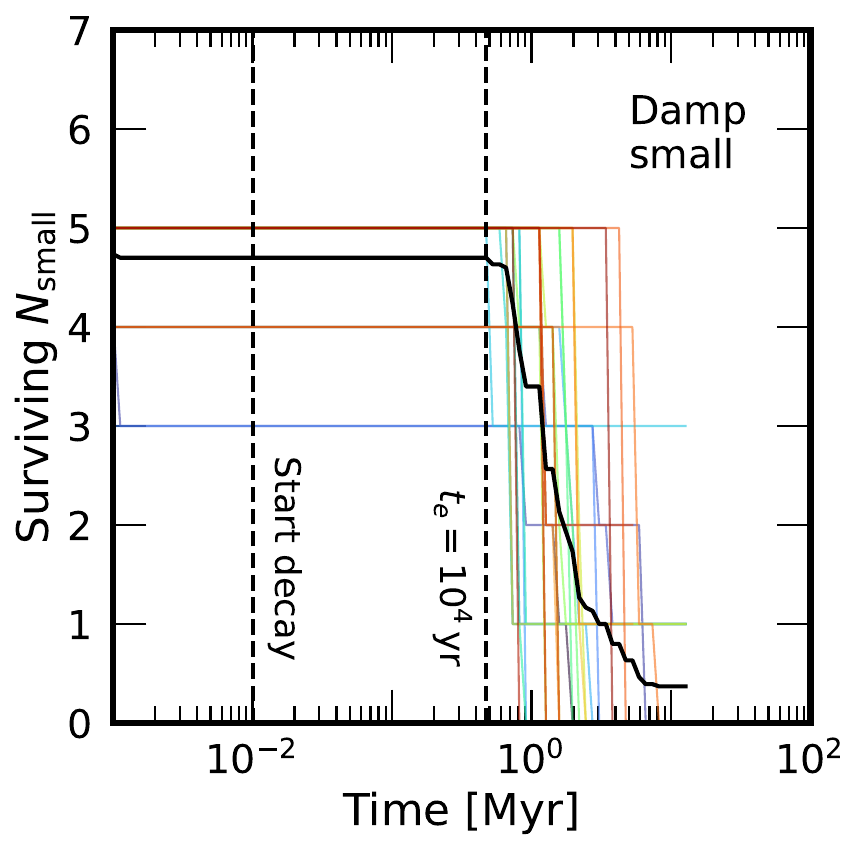}
\vspace{-0.5cm}
\caption{Similar to Fig.~\ref{fig:fidsmall}, but now assessing how damping the eccentricities of the small bodies affects their survival. At the start of the simulations, small bodies largely do not accrete because their eccentricities are quickly damped. The vertical dashed line marks the point when the exponential decay of damping has reached $t_e = 10^4$ yr. Around this point, the big bodies start to consume the small bodies and the system behaves similarly to the case without any damping.
\label{fig:damp_small}}
\end{figure}

In a second set of experiments, we only damp the small bodies. Our goal here is to understand whether damping might drive small bodies onto stable orbits that shield them from accretion even after the damping has disappeared. The damping prescription is the same as above, but applied to the small bodies instead. Figure \ref{fig:damp_small} shows that the small bodies are initially protected from accretion by the fast eccentricity damping. But once the damping rate has decayed enough that $t_e \sim t_{\rm acc} \sim 10^4$ yr (equation \ref{eqn:tacc}), the small bodies are consumed at a rate similar to the case without any damping.

\bsp	
\label{lastpage}
\end{document}